\documentclass[preprint,5p]{elsarticle}
\pdfoutput=1

\usepackage{amsfonts,amssymb,amsmath}	
\usepackage{graphicx}					
\usepackage{subfigure}					
\usepackage{hyperref}					
\usepackage[utf8]{inputenc}				

\graphicspath{{./figures/}}

\hypersetup{
  colorlinks	= true,	
  urlcolor		= blue,	
  linkcolor		= blue,	
  citecolor		= red	
}

\begin{document}
\begin{frontmatter}
\title{Fubini instantons in Dilatonic Einstein--Gauss--Bonnet theory of gravitation}

\author[1,2,3]{Bum-Hoon Lee}
\ead{bhl@sogang.ac.kr}
\author[2]{Wonwoo Lee}
\ead{warrior@sogang.ac.kr}
\author[1]{Daeho Ro}
\ead{daeho.ro@apctp.org}

\address[1]{Asia Pacific Center for Theoretical Physics, Pohang 37673, Korea}
\address[2]{Center for Quantum Spacetime, Sogang University, Seoul 04107, Korea}
\address[3]{Department of Physics, Sogang University, Seoul 04107, Korea}

\begin{abstract}
In this paper, we investigate various types of Fubini instantons in the Dilatonic Einstein--Gauss--Bonnet theory of gravitation which describes the decay of the vacuum state at a hilltop potential through tunneling without barrier. It is shown that the vacuum states are modified by the non-minimally coupled higher-curvature term. Accordingly, we present the new solutions which describe the tunneling from new vacuum states in anti-de Sitter and de Sitter backgrounds. The decay probabilities of the vacuum states are also influenced. We thus show that the semiclassical exponents can be decreased for specific parameter ranges, thereby increasing tunneling probability.
\end{abstract}

\begin{keyword}
Fubini instanton \sep DEGB theory \sep Tunneling without barrier
\PACS 04.62.+v \sep 98.80.Cq \\
Report Number: APCTP Pre2016-017
\end{keyword}

\end{frontmatter}


\section{\label{sec:1}Introduction} 
If an initial state of a potential is a false vacuum or metastable vacuum, the vacuum state can decay through tunneling \cite{Coleman:1977py,Kobzarev:1974cp,Goto:2016ofi} or rolling as a consequence of quantum and classical mechanics, respectively. The bounce solution describes the quantum tunneling process in Euclidean space, which has a negative mode \cite{Callan:1977pt}. The effect is more subtle in the presence of gravitation \cite{Koehn:2015hga,Lee:2014uza} The bounce solution determines the semi-classical exponent of the decay rate of an initial state. The decay rate per unit time per unit volume is given by $\Gamma/V = A e^{-B}$, where the pre-exponential factor $A$ is a functional determinant around the classical solution and the semi-classical exponent $B$ is the Euclidean action difference between the bounce solution and the background. The mechanism of the false vacuum decay is later developed with gravitation \cite{Coleman:1980aw,Parke:1982pm} and expanded upon in more detail \cite{Lee:2008hz}.

In the quantum tunneling process, a particle can penetrate a potential barrier to a region which is classically considered forbidden. The phenomenon of tunneling without barrier can also occur and it is very peculiar, because the tunneling occurs in a classically allowed region. If the time-scale of tunneling on the potential is shorter than that of rolling, the tunneling phenomenon can occure despite the lack of a potential barrier. This phenomenon was studied first as a Fubini instanton \cite{Fubini:1976jm,Lipatov:1976ny}, a bounce solution on the hilltop potential consisting of a negative quartic scalar field. The solution is used in the case of tunneling without a potential barrier between the top of a potential and an arbitrary state. The explicit form of the Fubini instanton was obtained through the conformal invariance of scalar field theory which allows the existence of a one-parameter family of a Euclidean solution of an arbitrary size but with the same probability. The conformal invariance can be broken by introducing a mass term. This makes the solution disappear \cite{Affleck:1980mp}. There have been intensive studies on tunneling without barrier in the absence of gravitation \cite{Lee:1985uv} and in the presence of gravitation \cite{Lee:1986saa} which are expanded upon later \cite{Kanno:2012zf,Lee:2012ug,Battarra:2013rba,Lee:2014ula}.

The inflationary universe scenario \cite{Guth:1980zm,Linde:1981mu,Albrecht:1982wi,Starobinsky:1980te} was proposed to solve a number of questions in the standard model of hot big-bang cosmology and it is now well established as a leading theory describing the early universe. Inflationary models with many different types of a potential have been studied and some of them have continued to stay relevant even after precision observation. Recently, it has been found that Planck data \cite{Ade:2015lrj} prefers hilltop and plateau inflationary models \cite{Kohri:2007gq,Martin:2013tda,Barenboim:2013wra,Coone:2015fha,Vennin:2015egh,Barenboim:2016mmw}. In addition, the instanton solutions in those potentials have renewed interest in the AdS/CFT correspondence \cite{deHaro:2006ymc,Papadimitriou:2007sj}. In line with these scenarios, it is worthwhile to observe quantum tunneling in a model with a hilltop potential as a case of Fubini instanton.

If one considers the very early universe of the Plank scale, the effects of the spacetime curvature become significant such that the early universe was in the quantum gravity regime. For this reason, we consider the higher-curvature Gauss--Bonnet (GB) term as a correction, in which the GB coefficient $\alpha$ is a dimensionless parameter. In four-dimensional spacetime, the theory using the GB term does not have a ghost particle or negative energy states \cite{Zwiebach:1985uq}. It also does not affect the equations of motion and the solutions. In order to introduce the contribution from the GB term, the term is coupled with a dilaton field with the coupling constant $\gamma$ which has a length dimension \cite{Kawai:1999pw,Guo:2010jr,Koh:2014bka}. The nonminimally coupled higher-curvature term appears in the first order $\alpha'$-correction (16$\alpha \kappa$ in the present paper) of the string effective action \cite{Boulware:1986dr}. This is called the Dilatonic Einstein--Gauss--Bonnet theory of gravitation (DEGB theory). It may provide a chance to avoid the initial singularity of the universe \cite{Antoniadis:1993jc,Rizos:1993rt,Easther:1996yd}. When taking into account the model of the Fubini instanton in the DEGB theory, the shape of the effective potential may change significantly depending on the signs of $\alpha$ and $\gamma$.

The paper is organized as follows: in the next section, we set up the basic framework with the action and the equations of motion for the DEGB theory. In section\ \ref{sec:3}, we analyze how the vacuum states are modified by the dilaton coupling with higher-order curvature terms. In section\ \ref{sec:4}, we present various types of solutions in DEGB theory. Finally in section\ \ref{sec:5}, we summarize our results and discuss their implication for cosmology.


\section{\label{sec:2}Set up} 
Let us consider the action that the scalar field is interacting with the Gauss--Bonnet (GB) term as follows:
\begin{multline} \label{eq:action}
S = \int_{\cal M} d^4x \sqrt{-g} \bigg[ \dfrac{R}{2\kappa} - \dfrac{1}{2} \partial_\mu \phi \partial^\mu \phi - U(\phi) + f(\phi) R_{\text{GB}}^2 \bigg]
\\ + S_{\text{YGH}},
\end{multline}
where $g = \det g_{\mu\nu}$ with the signs $(-,+,+,+)$, $\kappa = 8\pi G$, $R$ is the scalar curvature of a space-time manifold ${\cal M}$, $U(\phi)$ is a scalar field potential, and GB term is given by $R_{\text{GB}}^2 = R^{\mu\nu\rho\sigma}R_{\mu\nu\rho\sigma} - 4R^{\mu\nu}R_{\mu\nu} + R^2$. The function $f(\phi)$ is a coupling function between the scalar field and GB term. $S_{\text{YGH}}$ is the generalized York-Gibbons-Hawking boundary term \cite{York:1972sj,Gibbons:1976ue,Myers:1987yn,Brihaye:2008xu}.

By adopting the analytic continuation from Lorentzian to Euclidean, the action is changed with omitting boundary term as follows: 
\begin{equation} \label{eq:action.eu}
S_E = \int_{\cal M} d^4x \sqrt{g} \left[ \dfrac{-R}{2\kappa} + \dfrac{1}{2} \partial_\mu \phi \partial^\mu \phi + U(\phi) - f(\phi) R_{\text{GB}}^2 \right],
\end{equation}
where the signs of metric are now $(+,+,+,+)$ and the boundary term will be cancelled in the calculation of the difference between the action of the solution and background. The equation of motion for scalar field and Einstein's equation are
\begin{eqnarray} \label{eq:eom.phi}
0 &=& \nabla^2 \phi - U'(\phi) + f'(\phi) R_{\text{GB}}^2,
\\ \nonumber
0 &=& \dfrac{1}{2\kappa} \left( R_{\mu\nu} - \dfrac{1}{2} g_{\mu\nu} R \right) - \dfrac{1}{2} \partial_\mu \phi \partial_\nu \phi 
\\ \label{eq:eom.gr}
&& + \dfrac{1}{4} g_{\mu\nu} \partial_\rho \phi \partial^\rho \phi  + \dfrac{1}{2} g_{\mu\nu} U(\phi) + (\text{GB})_{\mu\nu},
\end{eqnarray}
where we use the primed notation for the derivative with respect to the scalar field $\phi$. The last term of second equation is obtained by GB term variation that in four-dimensional space is
\begin{multline} \label{eq:gb.mn}
(\text{GB})_{\mu\nu} = - 2 (\nabla_\mu \nabla_\nu f(\phi)) R + 2 g_{\mu\nu} (\nabla^2 f(\phi)) R 
\\ 
+ 4 (\nabla_\rho \nabla_\mu f(\phi)) R_\nu{}^\rho + 4 (\nabla_\rho \nabla_\nu f(\phi)) R_\mu{}^\rho 
\\ 
- 4 (\nabla^2 f(\phi)) R_{\mu\nu} - 4 g_{\mu\nu} (\nabla_\rho \nabla_\sigma f(\phi)) R^{\rho\sigma} 
\\
+ 4 (\nabla^\rho \nabla^\sigma f(\phi)) R_{\mu\rho\nu\sigma}.
\end{multline}
There were the terms linear in $f(\phi)$ but those terms are cancelled each other especially in four-dimensional space \cite{deWitt:1964}. Thus, $(\text{GB})_{\mu\nu}$ can be zero when the coupling function $f(\phi)$ is given by a constant which means that the scalar field and GB term are minimally coupled each other.

We consider Euclidean $O(4)$ symmetriy for the dominant contribution to the decay probability \cite{Coleman:1977th}. Then, the field $\phi$ as well as $\rho$ depends only on $\eta$ which is the radial coordinate of Euclidean space. The geometry is written as
\begin{equation} \label{eq:o4.metric}
ds^2 = d\eta^2 + \rho(\eta)^2 \big(d\chi^2 + \sin^2\chi (d\theta^2 + \sin^2\theta d\psi^2)\big).
\end{equation}
The scalar curvature and GB term turn out to be
\begin{equation} \label{eq:r.rgb}
R = - 6\dfrac{(\dot{\rho}^2 - 1 + \rho \ddot{\rho})}{\rho^2}, \quad \text{and} \quad R_{\text{GB}}^2 = 24\dfrac{\ddot{\rho}(\dot{\rho}^2 - 1)}{\rho^3},
\end{equation}
where the dotted notation denotes the derivative with respect to $\eta$. The equations of motion for $\phi$ and $(\eta,\eta)$, $(\chi, \chi)$ components of Einstein's equation are obtained from plugging Eq.\ \eqref{eq:o4.metric} into Eqs.\ \eqref{eq:eom.phi} and \eqref{eq:eom.gr} as follows:
\begin{align} \label{eq:eom2.phi}
0 =&\ \ddot{\phi} + 3\dfrac{\dot{\rho}}{\rho} \dot{\phi} - U'(\phi) + 24 f'(\phi)\dfrac{\ddot{\rho}(\dot{\rho}^2 - 1)}{\rho^3},
\\ \label{eq:eom2.gt}
0 =&\ \dfrac{3}{2\kappa}\dfrac{(\dot{\rho}^2 - 1)}{\rho^2} - \dfrac{1}{4} \dot{\phi}^2 + \dfrac{1}{2}U(\phi) - 12 \dot{f}(\phi) \dfrac{\dot{\rho} (\dot{\rho}^2 - 1)} {\rho^3}, 
\\ \nonumber
0 =&\ \dfrac{\dot{\rho}^2 - 1 + 2 \rho \ddot{\rho}}{2\kappa} + \dfrac{\rho^2}{4} \dot{\phi^2} + \dfrac{\rho^2}{2} U(\phi) - 8 \dot{f}(\phi) \dot{\rho}\ddot{\rho} 
\\ \label{eq:eom2.gtheta}
&\ - 4 \ddot{f}(\phi) (\dot{\rho}^2 - 1).
\end{align}
The last equation is different from the result obtained in Ref. \cite{Cai:2008ht}.

In order to solve the equations of motion, we should impose appropriate boundary conditions. In the presence of gravity, there need boundary conditions for not only $\phi$ but also $\rho$. These boundary conditions are divided into two types depending on the sign of $U(\phi_v)$. The maximum value of $\eta$ is $\eta_{\text{max}}=\infty$ for AdS and flat backgrounds, while it is finite for dS background which satisfies $\rho(\eta_{\text{max}})=0$.

For the flat and AdS space, we can impose the boundary conditions as follows:
\begin{equation}
\rho(0) = 0, \quad \dot{\rho}(0) = 1, \quad \dot{\phi}(0) = 0, \quad \text{and} \quad \phi(\eta_{\text{max}}) = \phi_v,
\end{equation}
where the first condition is for a geodesically complete space and second condition comes from the Eq.\ \eqref{eq:eom2.gtheta} and $\phi_v$ is the vacuum value of scalar field which was zero in Einstein's theory of gravitation (Einstein theory). For dS space, we can impose the boundary conditions as follows:
\begin{equation}
\rho(0) = 0, \ \ \rho(\eta_{\text{max}}) = 0, \ \ \dot{\phi}(0) = 0, \ \ \text{and} \ \ \dot{\phi}(\eta_{\text{max}})  = 0.
\end{equation}

We consider the potential
\begin{equation} \label{eq:po}
U(\phi) = - \dfrac{\lambda}{4} \phi^4 + U_0,
\end{equation}
where $\lambda$ is a positive constant \cite{Fubini:1976jm} and $U_0$ is a vacuum value of potential in Einstein theory. It plays a roll of cosmological constant as $\Lambda = \kappa U_0$ \cite{Lee:2012ug} while it is modified as $\Lambda=\kappa U(\phi_v)$ in DEGB theory. The coupling function is 
\begin{equation} \label{eq:cc}
f(\phi) = \alpha e^{-\gamma \phi},
\end{equation}
where $\alpha$ is the GB coefficient and $\gamma$ is the dilaton coupling constant. Here, we adopt the dimensionless parameters \cite{Lee:2012ug}. The change of variables still remain the equations of motion, Eqs.\ \eqref{eq:eom2.phi}, \eqref{eq:eom2.gt}, and \eqref{eq:eom2.gtheta}, same as before but the parameter $\lambda$ is disappeared. For numerical calculations we also set the initial value of $\eta$ by $0+\epsilon$ where $\epsilon \ll 1$ to avoid the initial divergence of equations of motion at $\eta=0$. Then, the initial values of $\phi$ and $\rho$ up to second order of $\epsilon$ are changed into
\begin{multline}
\phi(\epsilon) \approx \phi_0 - \dfrac{\epsilon^2}{8} \phi_0^3 + \cdots,  \qquad  \phi'(\epsilon) \approx - \dfrac{\epsilon}{4} \phi_0^3 + \cdots, \\
\rho(\epsilon) \approx \epsilon + \cdots, \hspace{24pt} \rho'(\epsilon) \approx 1 - \dfrac{\epsilon^2}{6} \kappa U(\phi_0) + \cdots.
\end{multline}


\section{\label{sec:3}Vacuum states}
We first determine the shape of the effective potential and vacuum states by solving the equations of motion with the given potential and GB coupling function in DEGB theory. This state is simply at $\phi=0$ in Einstein theory \cite{Lee:2014ula}. However, it becomes quite complicated depending on $\alpha$ and $\gamma$ in DEGB theory. For this reason, we analyze how the vacuum state can be modified by the dilaton coupling with higher-order curvature terms.

In order to find the value of $\phi_v$, the scalar field is supposed to initially at the vacuum state. Then, the scalar field stays at the vacuum state. In this regime, derivatives of the scalar field are vanished as follows:
\begin{equation} \label{eq:vac.con.1}
\dot{\phi}|_{\phi=\phi_v}  = 0, \qquad \text{and} \qquad \ddot{\phi}|_{\phi=\phi_v} = 0.
\end{equation}
By substituting this condition into Eqs.\ \eqref{eq:eom2.gt} and \eqref{eq:eom2.gtheta}, the analytic form of $\rho(\eta)$ is obtained for AdS and dS background as follows:
\begin{equation} \label{eq:vac.con.2}
\rho(\eta)  = \sqrt{\dfrac{3}{|\Lambda|}} \sinh \sqrt{\dfrac{|\Lambda|}{3}} \eta, \quad \text{and} \quad \rho(\eta)  = \sqrt{\dfrac{3}{\Lambda}} \sin \sqrt{\dfrac{\Lambda}{3}} \eta, 
\end{equation}
respectively. Recall that the cosmological constant is defined by $\Lambda = \kappa U(\phi_v)$. Then, the scalar field equation, Eq.\ \eqref{eq:eom2.phi}, can be simplified by using Eqs.\ \eqref{eq:vac.con.1} and \eqref{eq:vac.con.2} such as
\begin{equation} \label{eq:new}
0 = \phi_v^3 - \dfrac{8}{3} \alpha \gamma e^{-\gamma \phi_v} \Lambda^2.
\end{equation}
From this equation, the cosmological constant also satisfies $\Lambda = \pm \sqrt{3\phi_v^3/8\alpha\gamma e^{-\gamma \phi_v}}$, where the negative and positive sign correspond to the cases in the AdS and dS backgrounds, respectively. Finally, we can obtain the information about the vacuum state of scalar field by solving this equation with given parameters. For example, $\phi_v$ should be zero when $\alpha = 0$ which is the case of Einstein theory. If $\alpha \neq 0$, the equation cannot be solved analytically because it is higher-order and non-linear equation for $\phi_v$. Therefore, the values of $\phi_v$ are obtained numerically. In the numerical computation, we always choose the value of $\gamma$ to be positive for simplification of numerical results. Also, we fixed the value of $\kappa$ to be $0.1$ for all calculations.

\begin{figure*}[p]
\centering
\subfigure[\ Finding vacuum states]{\includegraphics[width=0.32\textwidth,height=0.15\textheight]{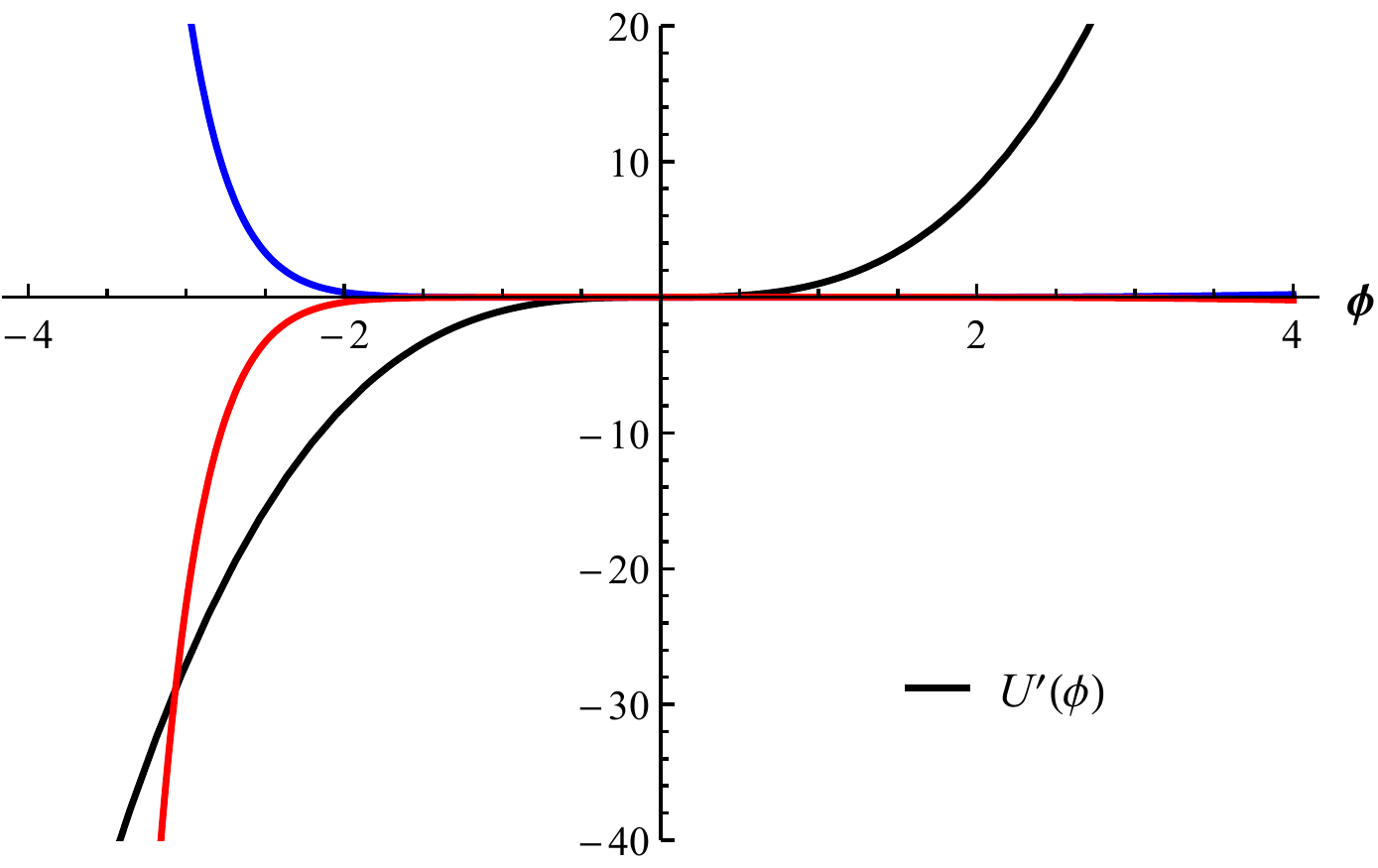}}
\subfigure[\ Sketch of effective potential for $\alpha < 0$]{\includegraphics[width=0.32\textwidth,height=0.15\textheight]{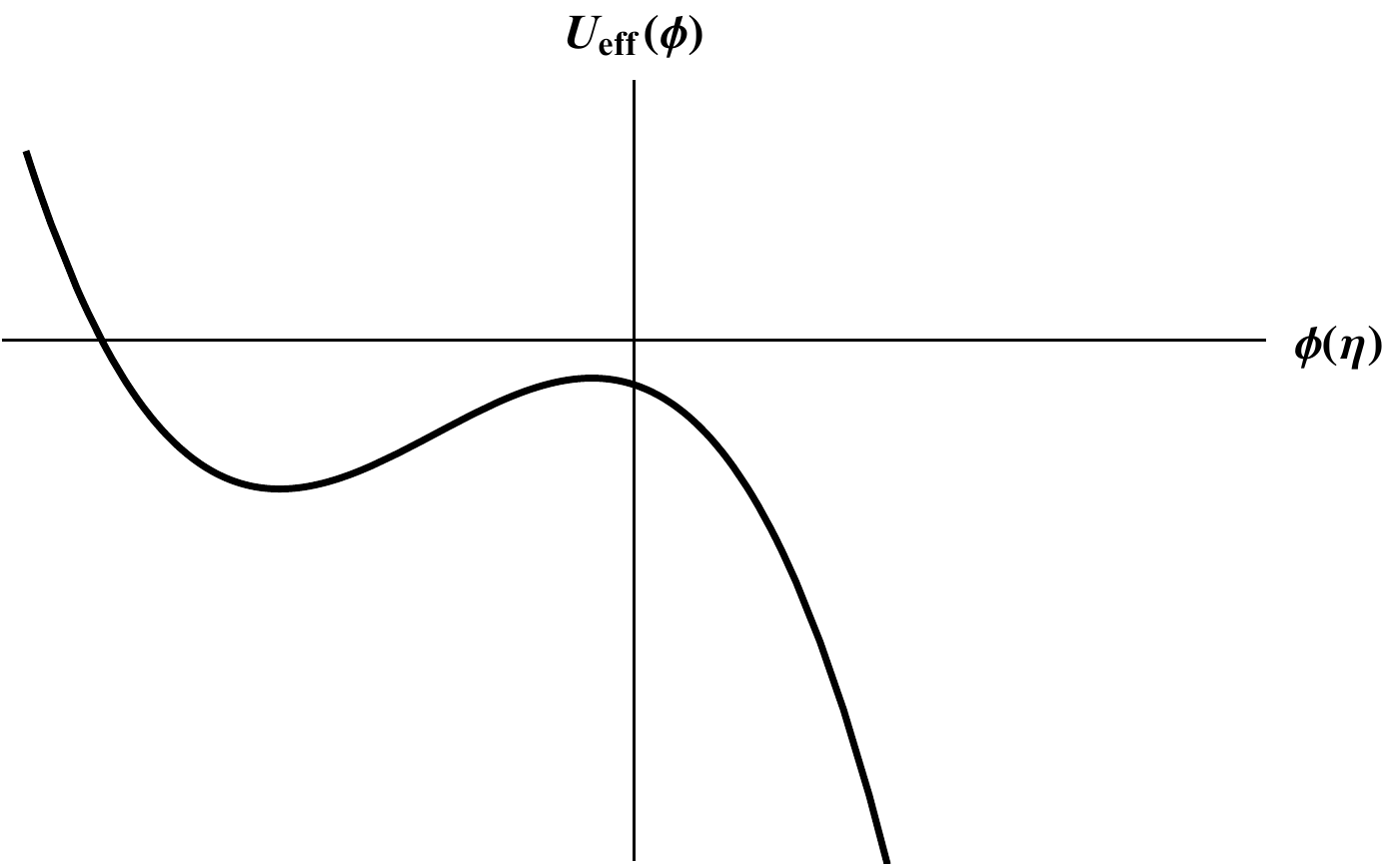}}
\subfigure[\ Sketch of effective potential for $\alpha > 0$]{\includegraphics[width=0.32\textwidth,height=0.15\textheight]{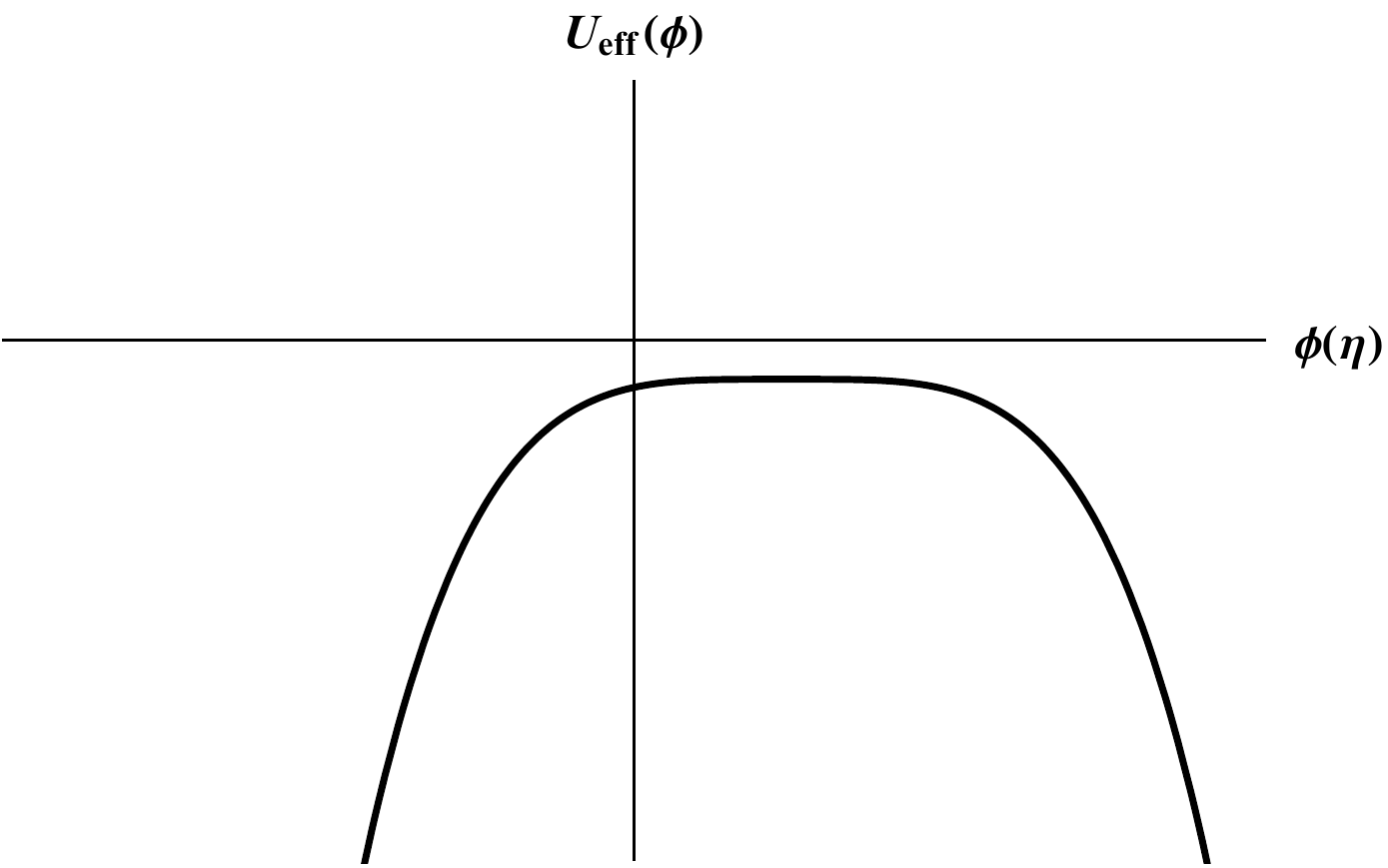}}
\caption{\label{fig:vac}(a) The red and blue line represent the first term in Eq.\ \eqref{eq:new} for the case of $\alpha < 0$ and $\alpha > 0$, and the black line represents the second term in Eq.\ \eqref{eq:new}. Since they cross each other twice and once for red and blue, respectively, there exist two vacuums for $\alpha < 0$ and one vacuum for $\alpha > 0$. Those are the static solutions in Eq.\ \eqref{eq:new}. The parameters are fixed as $\kappa = 0.1$, $\alpha = \pm 0.1$, $\gamma = 1.0$ and $U_0 = -0.3$. (b) Through the previous plot, we sketch the rough figure of the effective potential which is expected to have the form for $\alpha < 0$ and (c) for $\alpha > 0$ from the number of vacuum states, which means that it is not obtained from any equations but just guess. This is enough to deal with our motivation to find new types of a solution.}
\end{figure*}

Fig.\ \ref{fig:vac} represents the modified vacuum states in DEGB theory. For the negative values of $\alpha$, there exist two vacuum states and the effective potential is changed significantly, which makes the potential bounded below for the negative direction of the scalar field. However, for the positive values of $\alpha$, there exists only one vacuum state and the effective potential has the same form as the original hilltop potential. Note that the effective potential is not exactly obtained, but expected to be given form. It shows that the potential $U(\phi)$ is more affected by the GB term when $\alpha$ has different sign with $\gamma$.

\begin{figure*}[p]
\centering
\subfigure[\ $\alpha$ vs. $\phi_v$ with $\kappa=0.1$, $\gamma = 1.0$ and $U_0 = -0.3$]{\includegraphics[width=0.45\textwidth,height=0.15\textheight]{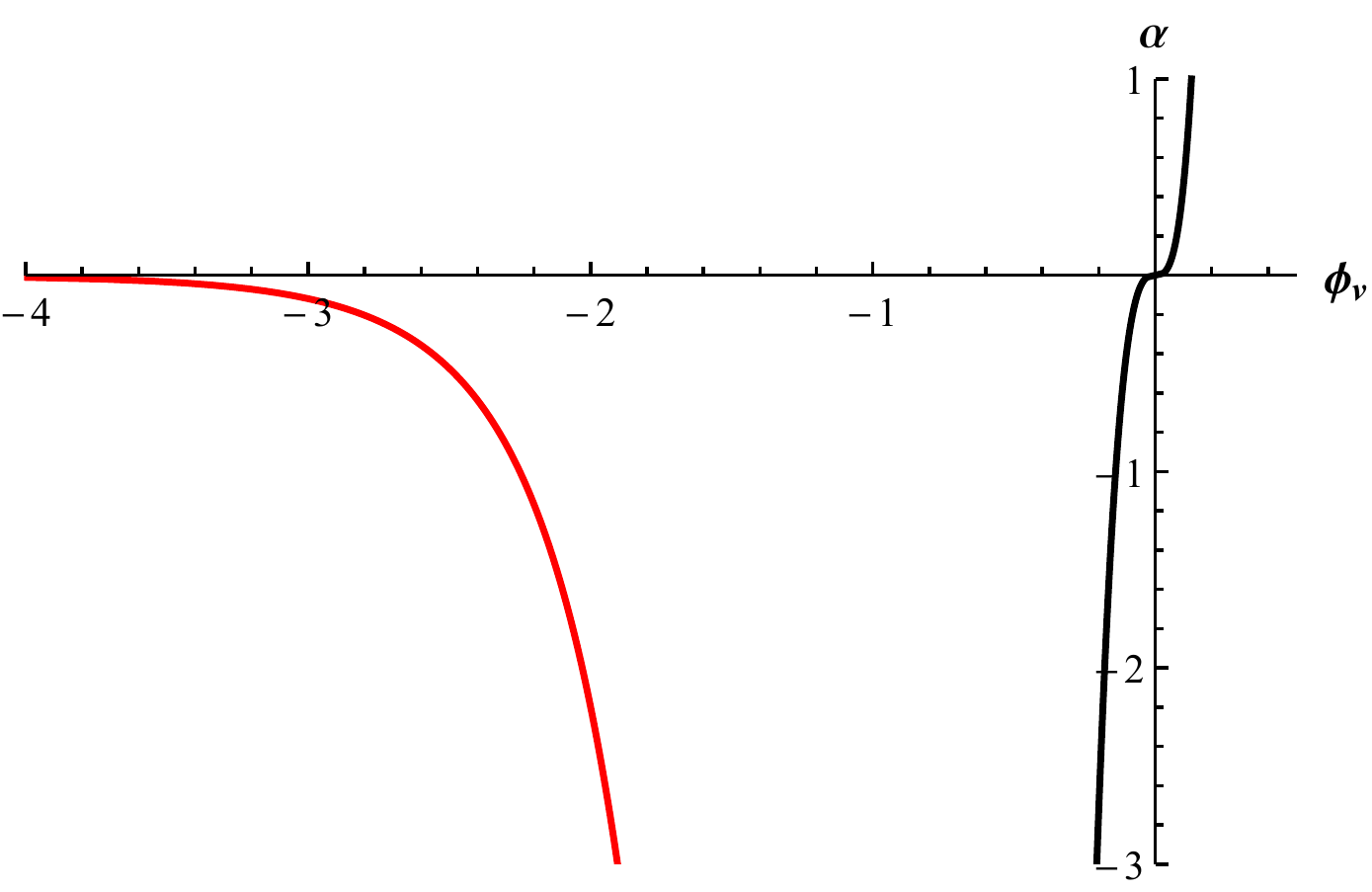}}
\hspace{0.03\textwidth}
\subfigure[\ $\alpha$ vs. $\phi_v$ with $\kappa=0.1$, $\gamma = 1.0$ and $U_0 = 0.3$]{\includegraphics[width=0.45\textwidth,height=0.15\textheight]{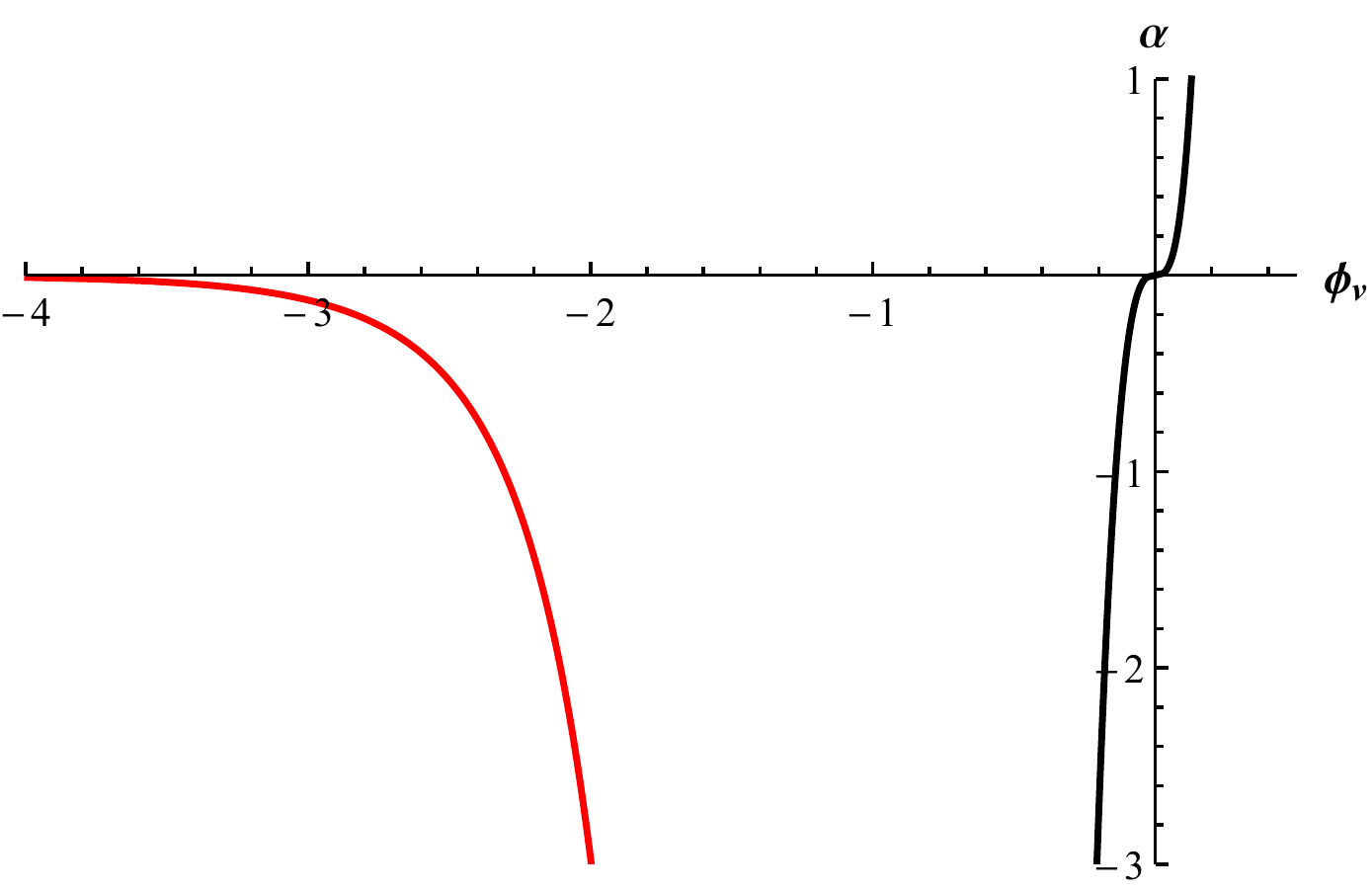}}
\\
\subfigure[\ $\alpha$ vs. $\phi_v$ with $\kappa=0.1$, $\gamma = 8.0$ and $U_0 = -0.3$]{\includegraphics[width=0.45\textwidth,height=0.15\textheight]{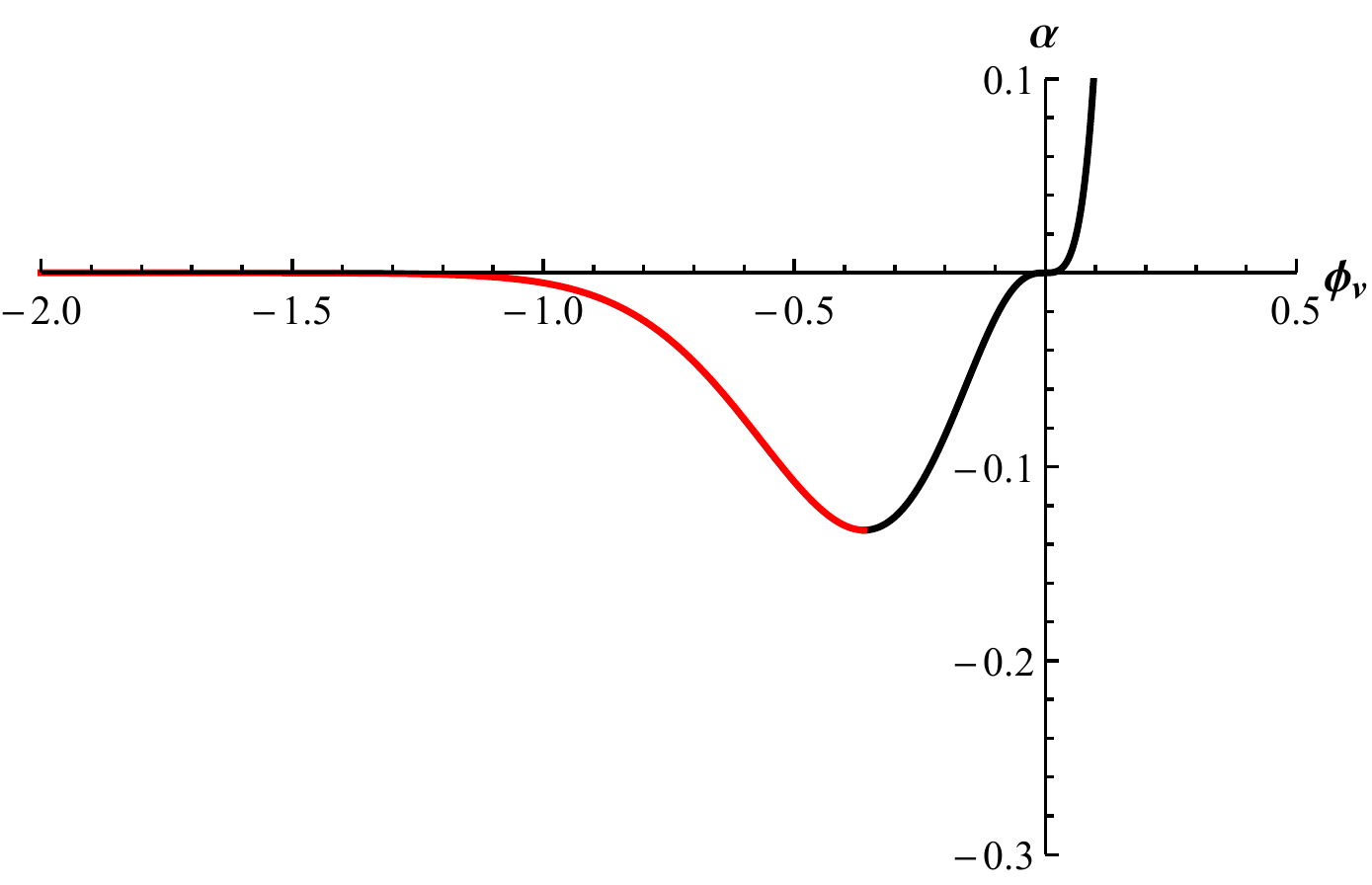}}
\hspace{0.03\textwidth}
\subfigure[\ $\alpha$ vs. $\phi_v$ with $\kappa=0.1$, $\gamma = 8.0$ and $U_0 = 0.3$]{\includegraphics[width=0.45\textwidth,height=0.15\textheight]{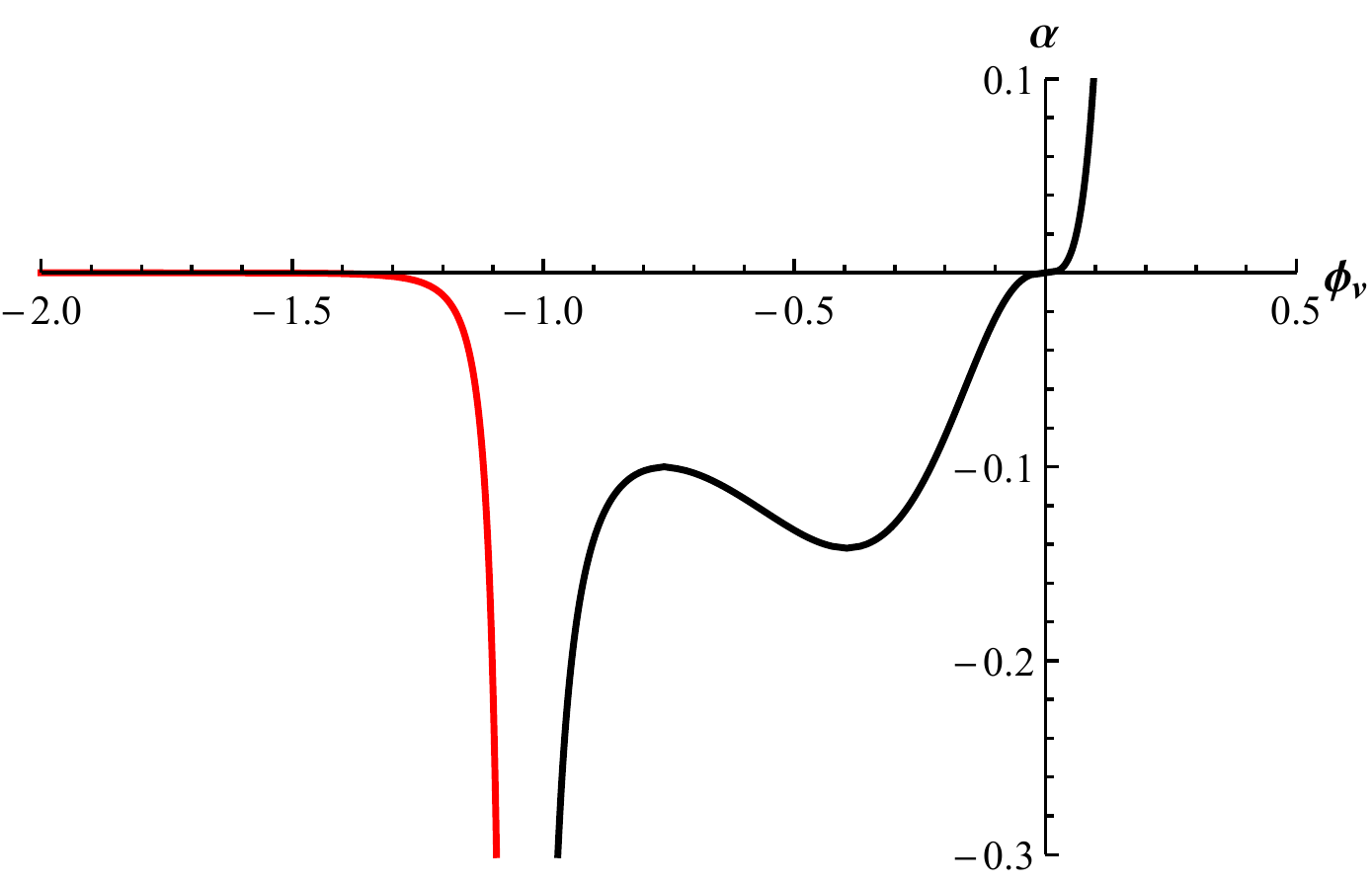}}
\caption{\label{fig:alpha.vac}(a) The black and red line represent the vacuum states which are obtained from the simplified scalar field equation. The red line is appeared only for $\alpha < 0$. This plot shows the values of vacuum state with respect to $\alpha$ with fixed parameter $\gamma=1.0$ for AdS background and (b) for dS background. (c) When $\gamma=8.0$, it is shown that the number of vacuums become one at specific negative $\alpha$ and disappear when $\alpha$ further decreases for AdS background and (d) the number of vacuums increases at specific range of negative $\alpha$ and decreases again when $\alpha$ further decreases for dS background.}
\end{figure*}

Fig.\ \ref{fig:alpha.vac} represents the value of $\phi_v$ with respect to $\alpha$ for AdS and dS backgrounds. First two plots look very similar, but the exact values of $\phi_v$ are different. For example, the first and second vacuum states are at $\phi_v = -0.0634741$ and $\phi_v = -3.0678$ for AdS background, but those for dS background at $\phi_v = -0.0634729$ and $\phi_v = -3.08818$ when $\alpha = -0.1$. In this case, the black line indicates the change of vacuum states and the magnitude depend on the value of $\alpha$. The red line only appears in the region of negative $\alpha$. Both lines are becoming closer when $\alpha$ decreases and expected to meet at some point. It can be seen by increasing $\gamma$ as the other two plots. Two lines eventually meet at the specific negative value of $\alpha$ and there is no vacuum state when $\alpha$ further decreases in the AdS background. However in the dS background, the number of vacuum states increases from one up to four with the specific range of $\alpha$ which makes the effective potential be more complicated. Both lines in the dS background are also expected to meet at some point by decreasing $\alpha$.

\begin{figure*}[p]
\centering
\subfigure[\ AdS background case]{\includegraphics[width=0.45\textwidth,height=0.16\textwidth]{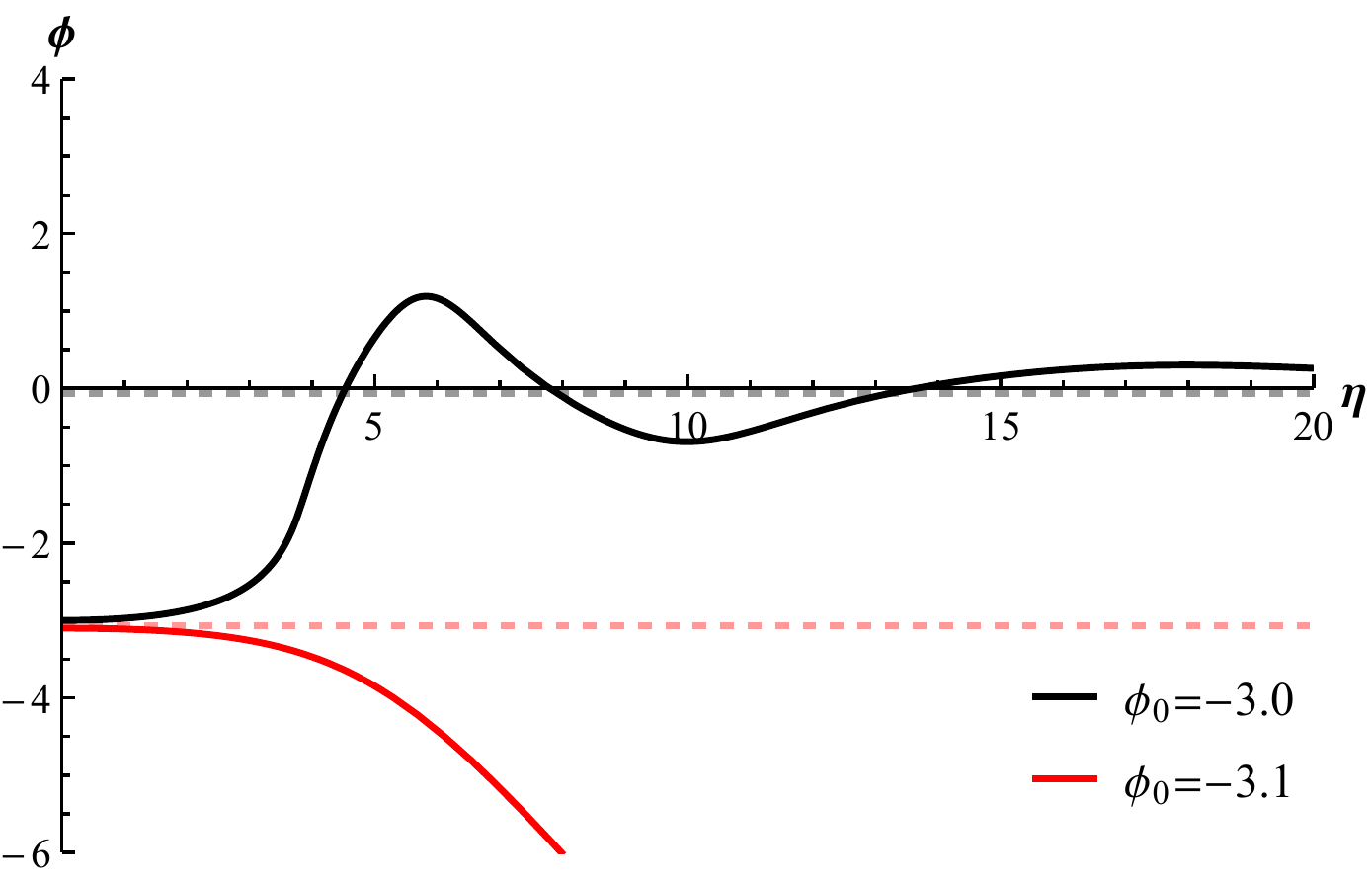}}
\hspace{0.03\textwidth}
\subfigure[\ dS background case]{\includegraphics[width=0.45\textwidth,height=0.16\textwidth]{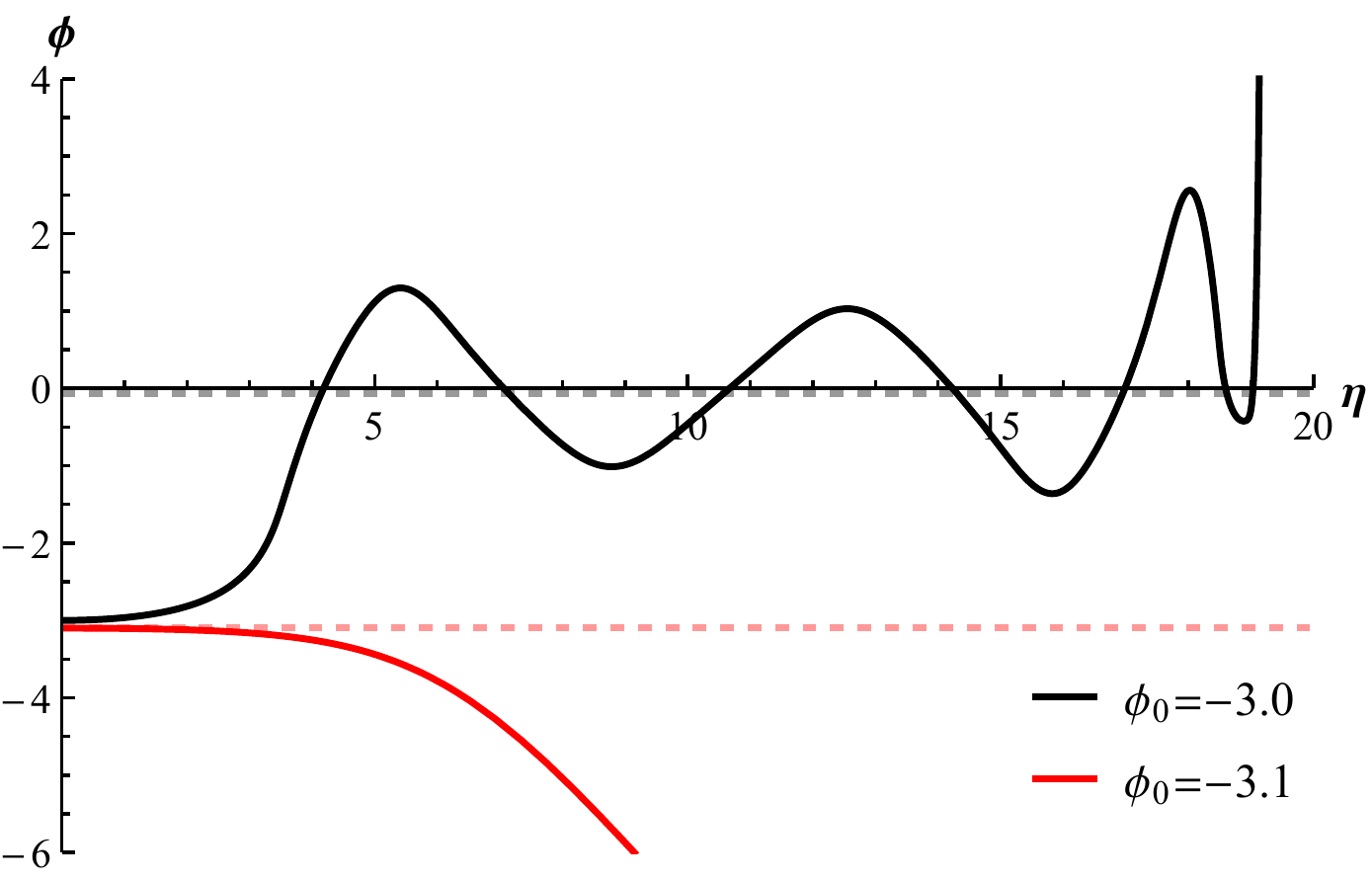}}
\caption{\label{fig:new.vac}Scalar field behavior with the initial values near the second $\phi_v$ with $\kappa=0.1$, $\alpha = -0.1$, $\gamma=1.0$ and $U_0=\mp0.3$. It converges to or oscillates near the other vacuum when the initial value of scalar field $\phi_0$ is larger than $\phi_v$, similarly for the case with Einstein theory. However, it grows to negative infinity and cannot satisfy the boundary condition when $\phi_0$ is smaller than $\phi_v$.}
\end{figure*}

Fig.\ \ref{fig:new.vac} shows the limitation of the initial scalar field values with $\alpha=-0.1$. This figure represents the evolution of scalar fields with specific initial values. When the scalar fields have more higher values than the second vacuum state, they converge to the first vacuum state in the AdS background and oscillate near the first vacuum state in the dS background. However, the scalar fields grow into the negative infinity when they initially have the lower value than the second vacuum state in both AdS and dS background. Then, they cannot satisfy the boundary condition.


\section{\label{sec:4}Numerical solutions and decay rates}
In this section, we present the solutions by solving the equations of motion in the Ads and dS backgrounds. We first briefly review the results of the solutions in Einstein theory \cite{Lee:2014ula}. There are three parameters to determine a solution, $\kappa$, $\phi_0$ and $U_0$. The crossing number of the top (or the vacuum state) in the inverted potential is the number of oscillations \cite{Hackworth:2004xb,Lee:2011ms,Battarra:2012vu}. In the AdS background, any set of parameters allows a oscillating solution with different number of oscillations. The solutions with the specific number of oscillations have the maximum or minimum initial value of scalar field, we call it the marginal solution. In general, all the solutions have infinite exponent $B$ while the marginal solutions have finite values. In the dS background, the specific set of parameters allows a oscillating solution with the different number of oscillations and the equations of motion diverge in other cases. There exist two types of solutions which are $Z_2$-symmetric and $Z_2$-asymmetric. When the solution starts at $\phi_0$ and  ends at $\pm\phi_0$, then it is called $Z_2$-symmetric. For $Z_2$-asymmetric, the solution starts at $\phi_1$ or $\phi_2$ and ends at $\phi_2$ or $\phi_1$ where $\phi_1 \neq \phi_2$.

If we plot all the data with respect to $\alpha$, it is numerous. Thus, we select two solutions in the AdS background which are the solutions with $\phi_0 = \pm 1.5$ and one solution in the dS background which corresponds to the $Z_2$-symmetric solution in Einstein theory. In this paper, we concentrate on how those solutions are modified by the dilaton coupling with higher-order curvature terms.

\begin{figure}[!b]
\centering
\subfigure[\ Solutions initially starts from $\phi_0=-1.5$]{\includegraphics[width=0.45\textwidth,height=0.145\textheight]{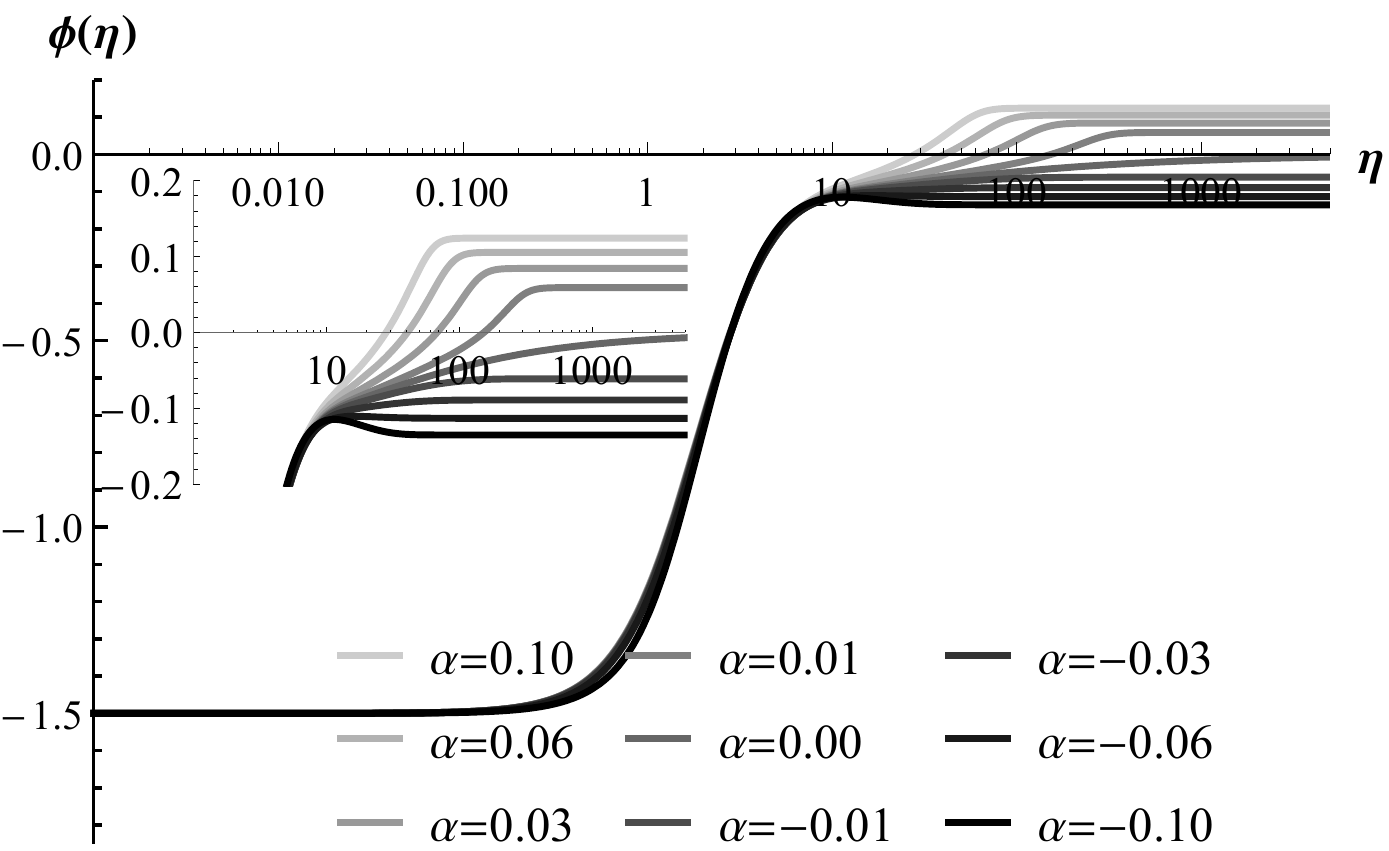}}
\\
\subfigure[\ Solutions initially starts from $\phi_0=1.5$]{\includegraphics[width=0.45\textwidth,height=0.145\textheight]{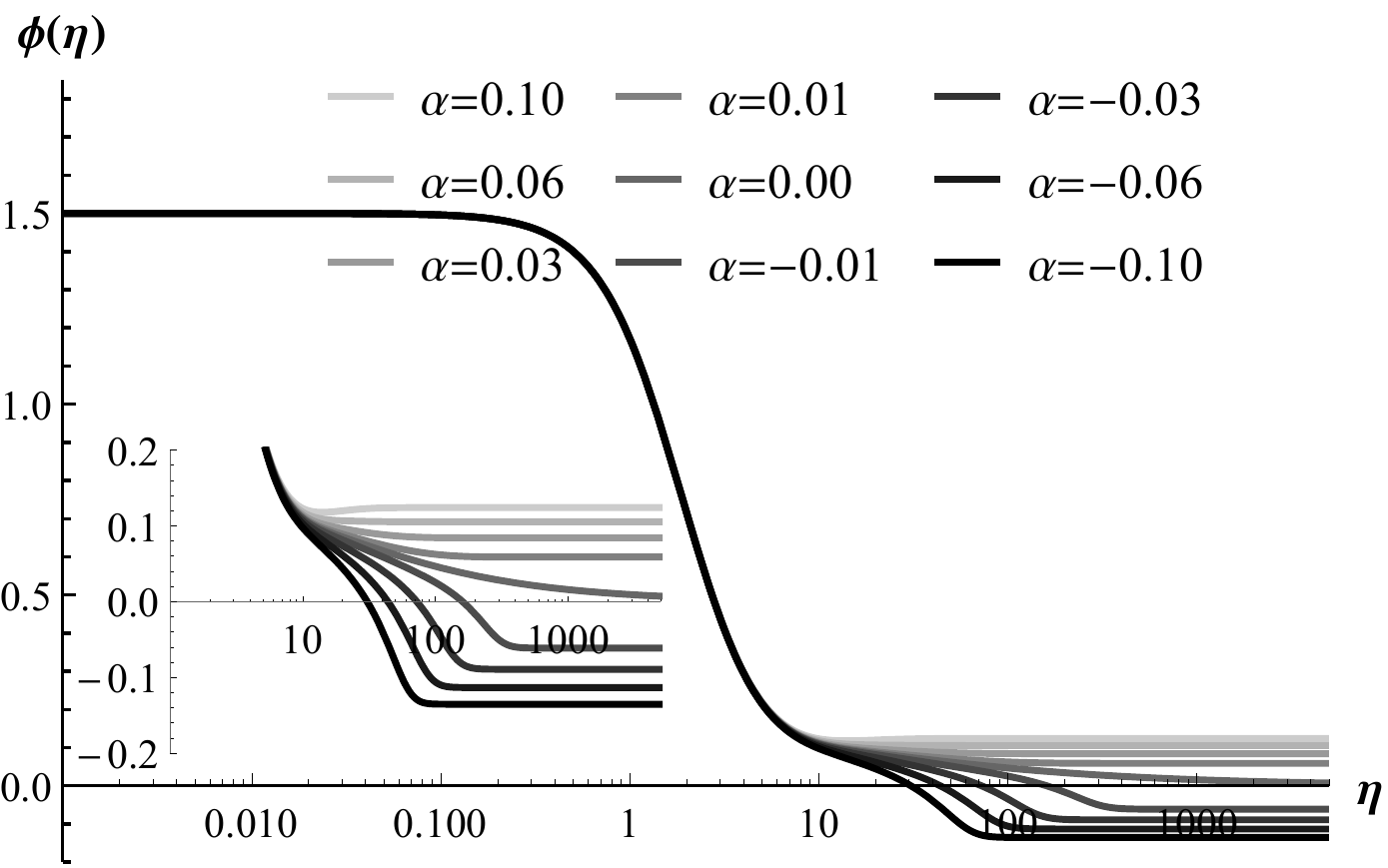}}
\caption{\label{fig:ads.alpha}Solutions of $\phi(\eta)$ vs. $\eta$ in the AdS background with respect to $\alpha$. The parameters are fixed as $\kappa=0.1$, $\gamma=1.0$ and $U_0=-0.3$.}
\end{figure}

\begin{figure}[!b]
\centering
\subfigure[\ Solutions initially starts from $\phi_0=-1.5$]{\includegraphics[width=0.45\textwidth,height=0.145\textheight]{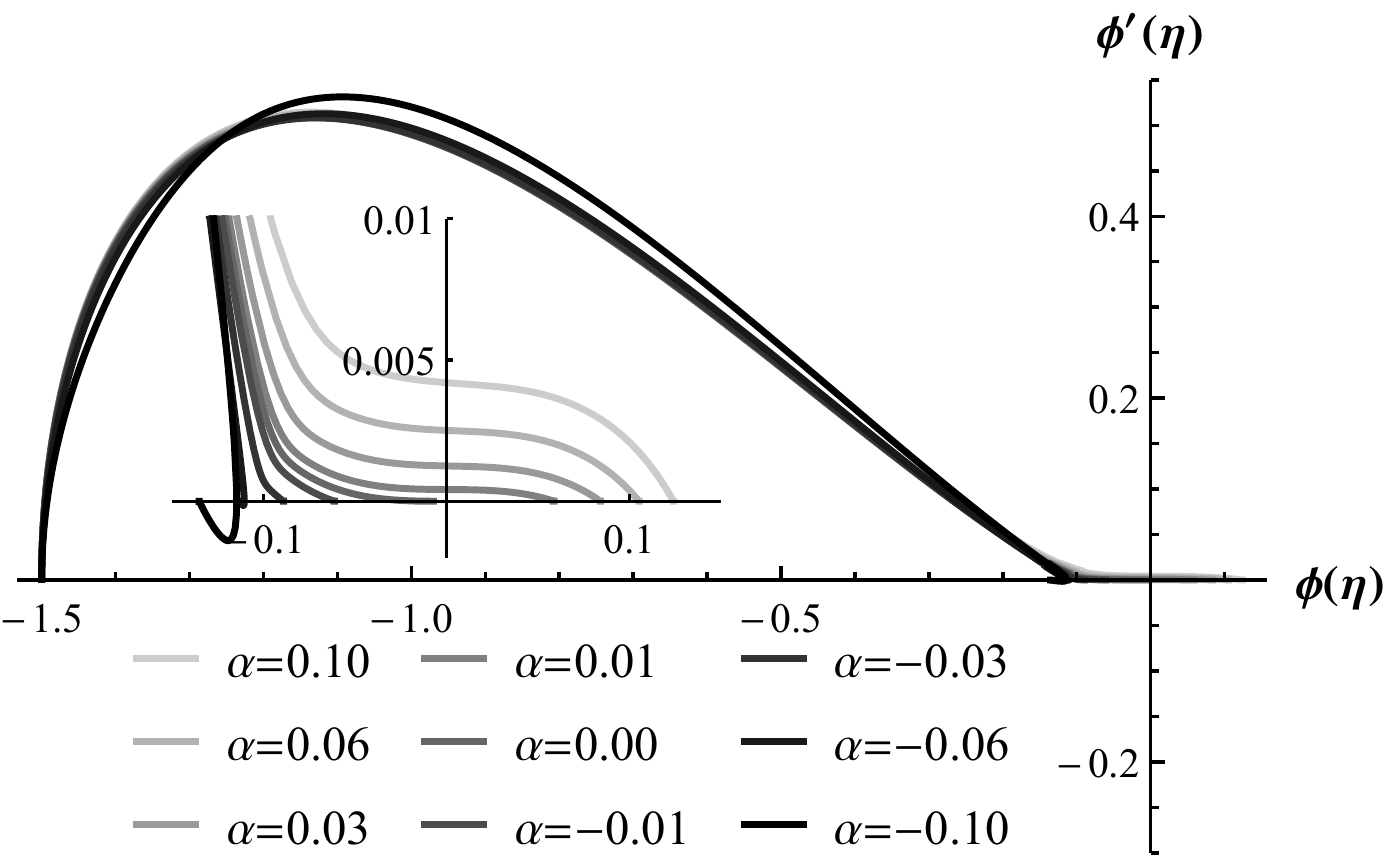}}
\\
\subfigure[\ Solutions initially starts from $\phi_0=1.5$]{\includegraphics[width=0.45\textwidth,height=0.145\textheight]{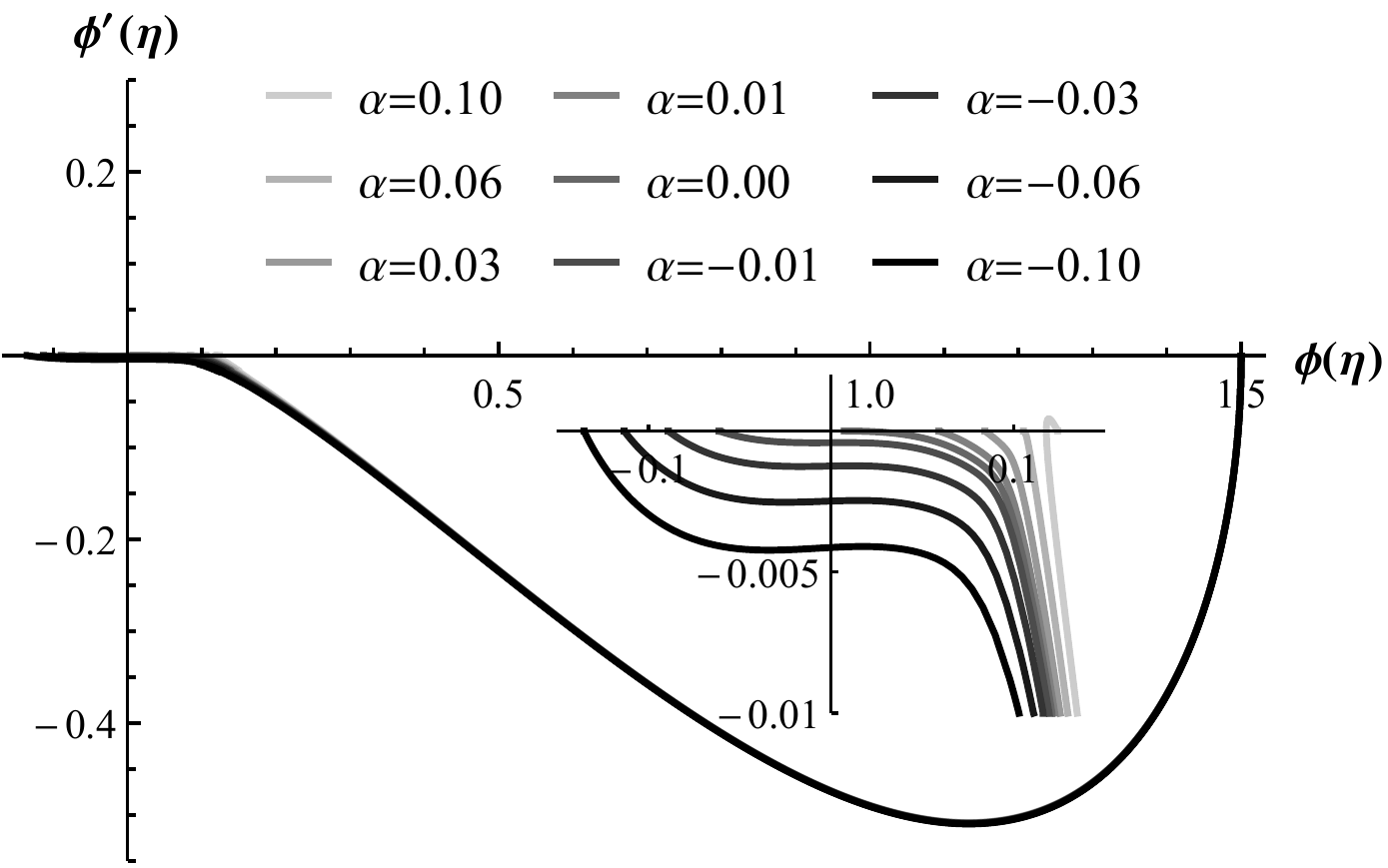}}
\caption{\label{fig:ads.alpha2}Solutions of $\phi'(\eta)$ vs. $\phi(\eta)$ in the AdS background with respect to $\alpha$. The parameters are fixed as $\kappa=0.1$, $\gamma=1.0$ and $U_0=-0.3$.}
\end{figure}

Figs.\ \ref{fig:ads.alpha} and \ref{fig:ads.alpha2} represent two selected solutions in the AdS background with respect to $\alpha$. The gray line of a solution becomes lighten or darken when $\alpha$ increases or decreases, respectively. The solutions converge into the vacuum states which are determined by the values of $\alpha$. The sign of the first vacuum state, $\phi_v$ follows that of $\alpha$ and the magnitude of $\phi_v$ increases when the absolute value of $\alpha$ increases. In order to see the details near the vacuum state, small inset figures are added in each figure. Fig.\ \ref{fig:ads.alpha} shows the solutions of $\phi(\eta)$ with respect to $\eta$ and Fig.\ \ref{fig:ads.alpha2} shows the solutions of $\phi'(\eta)$ with respect to $\phi(\eta)$. Most of the solutions do not have oscillatory behavior. However, the solutions only with $\phi_0=\pm 1.5$ and $\alpha=\mp 0.10$ have one oscillation. This fact can be easily recognized by the inset plots in Fig.\ \ref{fig:ads.alpha2}. 

All the solutions of scalar fields are behaving very similar until $\eta \sim 10$ and separating after that into there own vacuums. Therefore, the magnitude of vacuum value determines how many times the solution oscillate for fixed $\phi_0$, in other words, it is determined by $\alpha$. Increase of $|\alpha|$ which has the same sign with $\phi_0$ thus enhance those oscillatory behavior. However, it can be suppressed if $\alpha$ and $\phi_0$ have different sign. In addition, the oscillatory behavior is more enhanced when $\alpha$ has the negative value than positive, which can be seen by comparing the darkest line of Fig.\ \ref{fig:ads.alpha2}(a) and lightest line of Fig.\ \ref{fig:ads.alpha2}(b). The difference comes from the existence of the second vacuum state which is only appeared in the region of the negative $\alpha$. The second vacuum state restricts the selection of $\phi_0$ and leads to significant change of the effective potential. It is expected that the second vacuum state highly affects the oscillatory behavior.

\begin{table*}[!ht]
\centering
\begin{tabular}{|c||c|c|c|c|c|c|c|c|c|} \hline
$\alpha$ & $-0.10$ & $-0.06$ & $-0.03$ & $-0.01$ & $0$ & $0.01$ & $0.03$ & $0.06$ & $0.10$ \\ \hline
$\phi_v$ & $-0.06347$ & $-0.05336$ & $-0.04219$ & $-0.02913$ & $0$ & $0.02857$ & $0.04104$ & $0.05152$ & $0.06090$ \\ \hline
$\phi_0$ & $5.2629$ & $5.0842$ & $4.8807$ & $4.6274$ & $3.9355$ & $2.8734$ & $2.1955$ & $1.5536$ & $1.0391$ \\ \hline
$B$ & $32.7$ & $32.9$ & $32.7$ & $32.0$ & $29.8$ & $27.2$ & $26.1$ & $26.1$ & $29.1$ \\ \hline
\end{tabular}
\caption{\label{tab:1}Exponent $B$ of decay rate for the marginal solutions in the AdS background with respect to $\alpha$.}
\end{table*}

The solutions with arbitrary parameters in the AdS background have infinite exponent $B$ as we discussed earlier. Thus, we find the marginal solutions which does not oscillate with respect to $\alpha$. Table\ \ref{tab:1} represents the exponent $B$ of decay rate for those marginal solutions. The exponent is defined by $B = S_{\text{bs}} - S_{\text{bg}}$ where $S_{\text{bs}}$ and $S_{\text{bg}}$ are Euclidean action for the bounce solution and background, respectively. The exponent $B$ turns out to be
\begin{multline}
B = 2\pi^2 \int_0^{\rho_{\text{m}}} d\rho \dfrac{\rho^3}{\dot{\rho}} 
\\
\bigg( S_E|_{\eta\rightarrow \rho^{-1}} - S_E|_{\eta\rightarrow \sqrt{\frac{3}{|\Lambda|}}\sinh^{-1}\sqrt{\frac{|\Lambda|}{3}}\rho, \phi\rightarrow\phi_v}\bigg)
\label{eq:bounce}
\end{multline}
where $\rho_{\text{m}}$ is the value of the maximum $\rho$ which is infinite, in general. However, for the marginal solution, it can be finite because the integration values after $\rho_{\text{m}}$ are not significant. 

The exponent $B$ and $\phi_0$ increase when $\alpha$ decreases in the vicinity of $\alpha = 0$. The increase of the difference between $\phi_0$ and $\phi_v$ is expected to give more higher value during the integration of exponent $B$. However, the exponent $B$ decreases when $\alpha$ decreases further. Similarly, the exponent $B$ and $\phi_0$ decrease when $\alpha$ increases, but those values increase when $\alpha$ increases further. Since the decay rate is proportional to $e^{-B}$, the small exponent value means the case is more probable. Thus, the case with the small positive or large negative values of $\alpha$ is dominant.

\begin{figure}[!b]
\centering
\subfigure[\ Plot of $\phi(\eta)$ vs. $\eta$]{\includegraphics[width=0.45\textwidth,height=0.145\textheight]{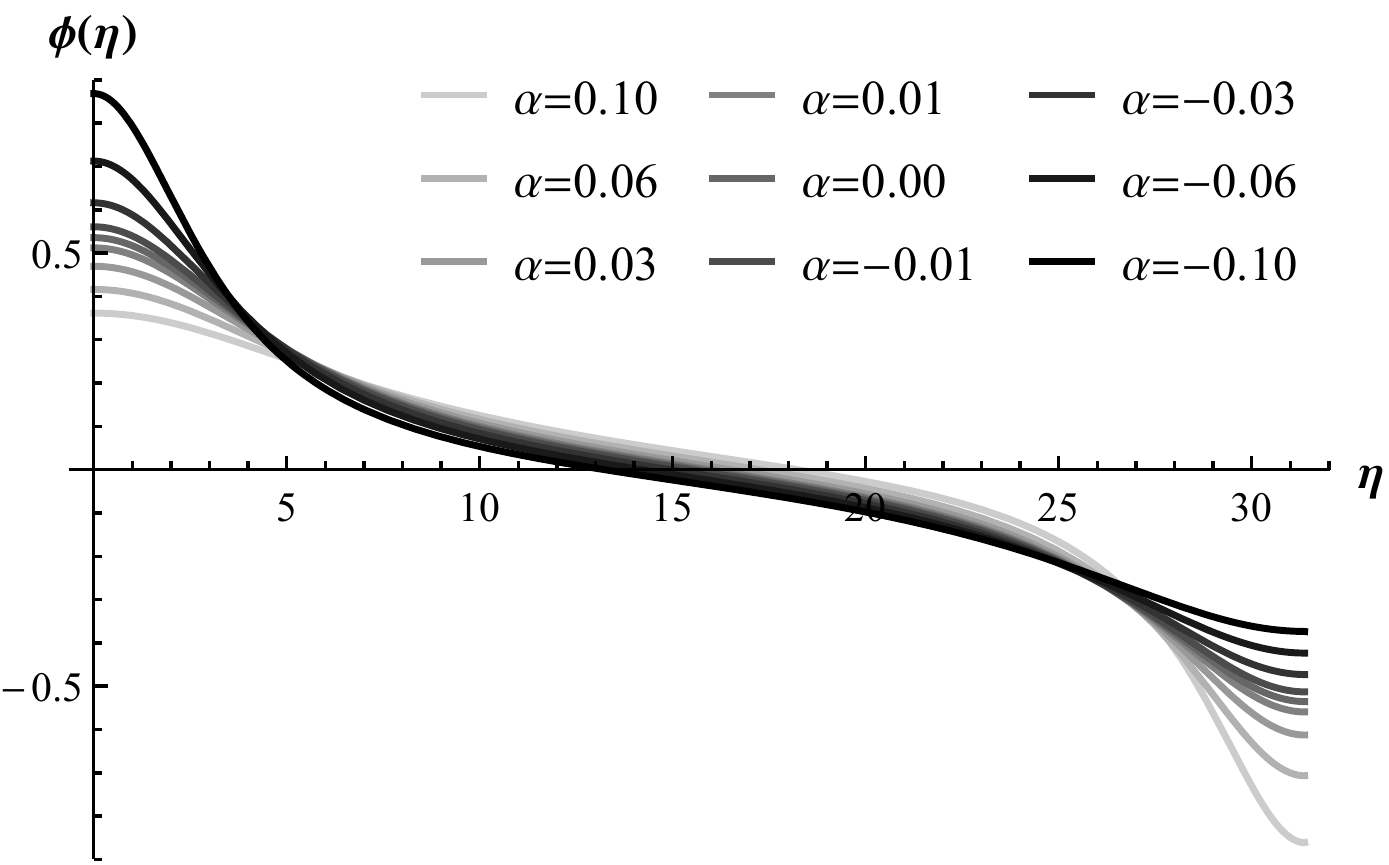}}
\hspace{0.03\textwidth}
\subfigure[\ Plot of $\phi'(\eta)$ vs. $\phi(\eta)$]{\includegraphics[width=0.45\textwidth,height=0.145\textheight]{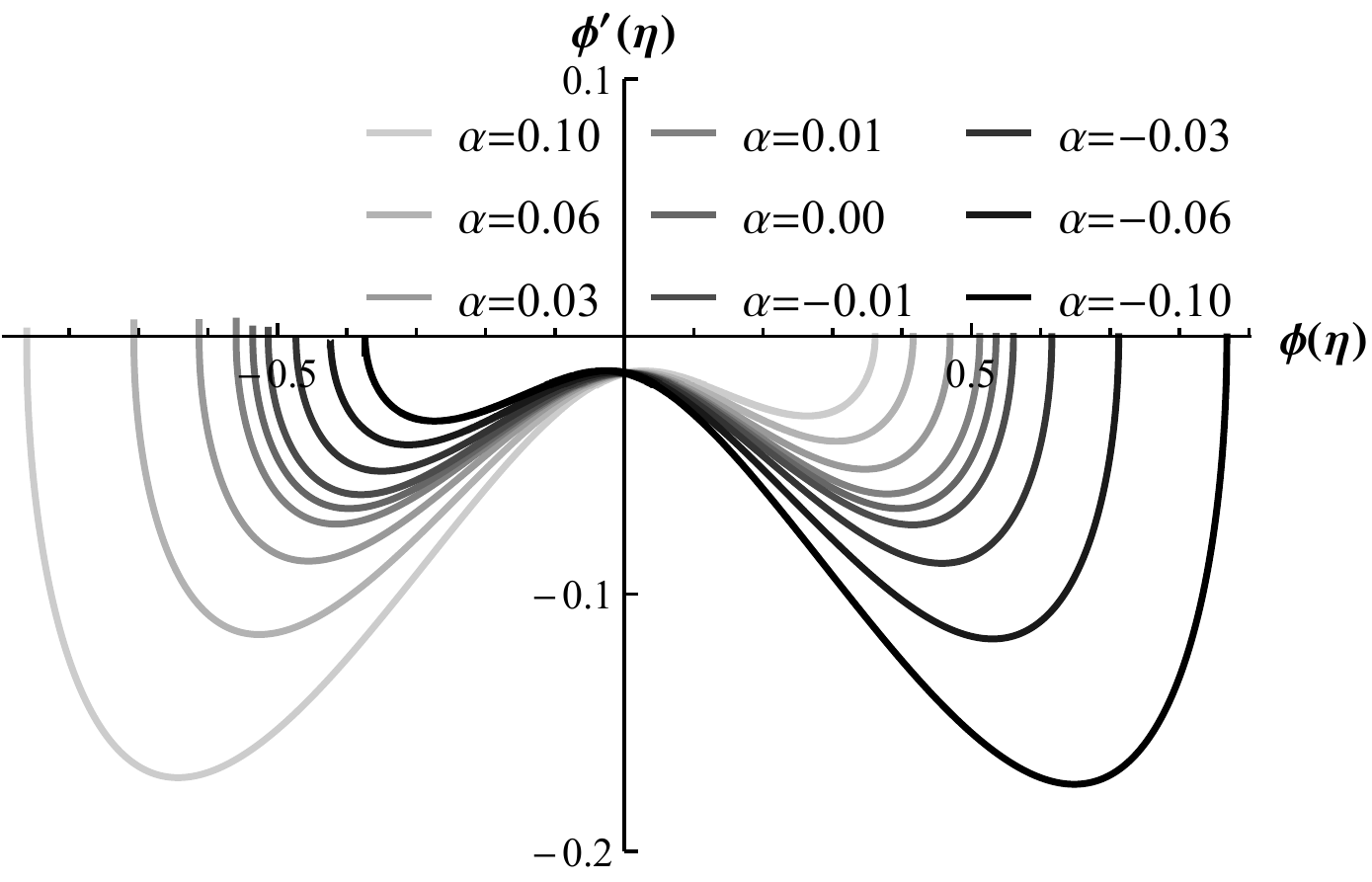}}
\caption{\label{fig:ds.alpha.phi}Selected type of solutions in the dS background with respect to $\alpha$. The parameters are fixed as $\kappa=0.1$, $\gamma=1.0$ and $U_0=0.3$.}
\end{figure}

Fig.\ \ref{fig:ds.alpha.phi} represent a selected solution in the dS background with respect to $\alpha$. It is $Z_2$-symmetric one as the case for $\alpha=0$ in Einstein theory, however, GB term brake $Z_2$-symmetry. The gray line of solution becomes lighten or darken when $\alpha$ increases or decreases, the same as the AdS solution. In Fig.\ \ref{fig:ds.alpha.phi}(a), the initial (or final) value of the scalar field increase (or decreases) when $\alpha$ decreases. In Fig.\ \ref{fig:ds.alpha.phi}(b), the absolute value of $\phi'(\eta)$ increases when $|\alpha|$ increases. The solutions with $\alpha = \pm0.10$ look symmetric under the inversion of y-axis, but there exists a little difference. The difference is enhanced by increasing $\gamma$.

\begin{figure}[!b]
\centering
\subfigure[\ For AdS background]{\includegraphics[width=0.45\textwidth,height=0.145\textheight]{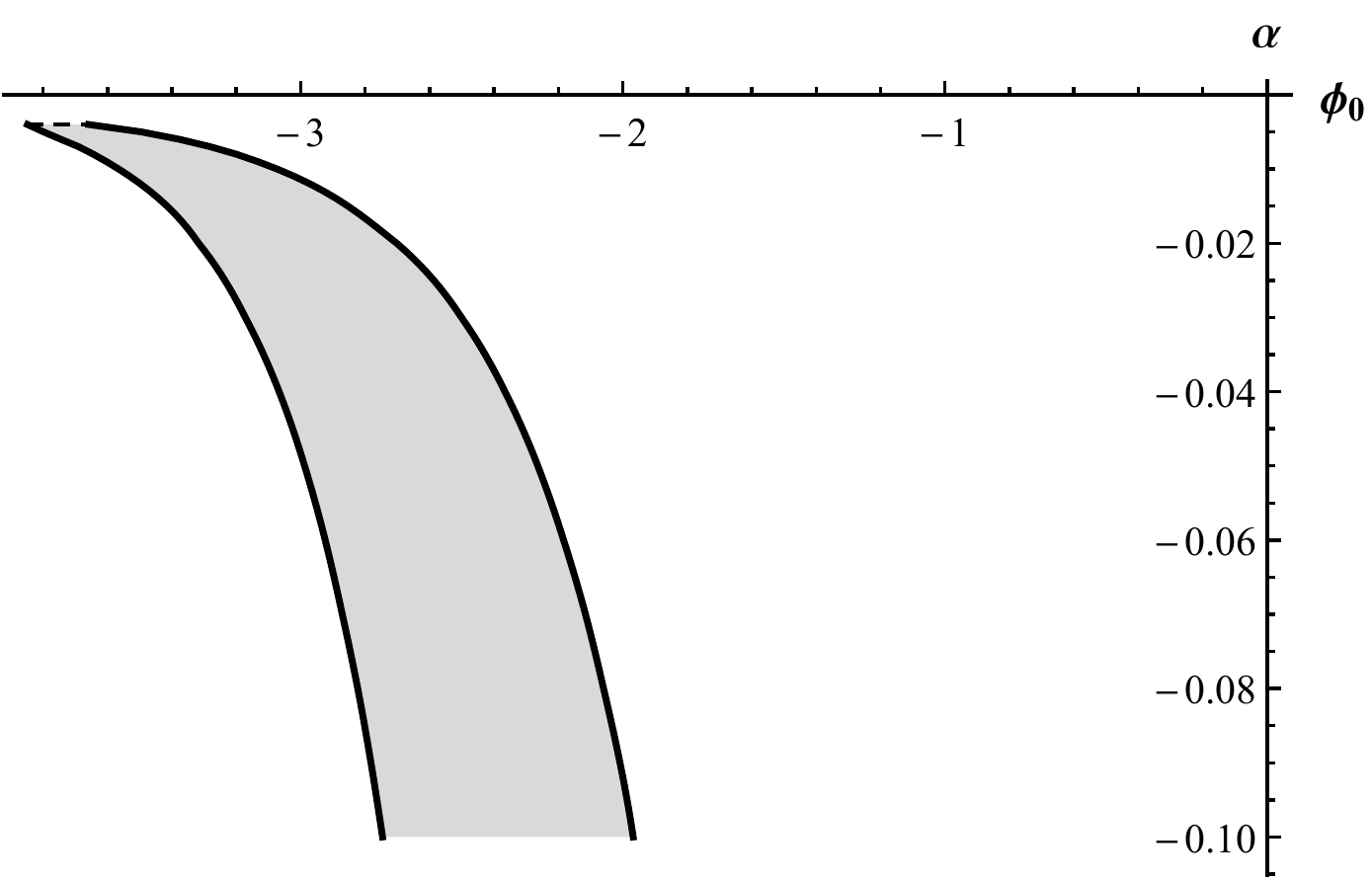}}
\hspace{0.03\textwidth}
\subfigure[\ For dS background]{\includegraphics[width=0.45\textwidth,height=0.145\textheight]{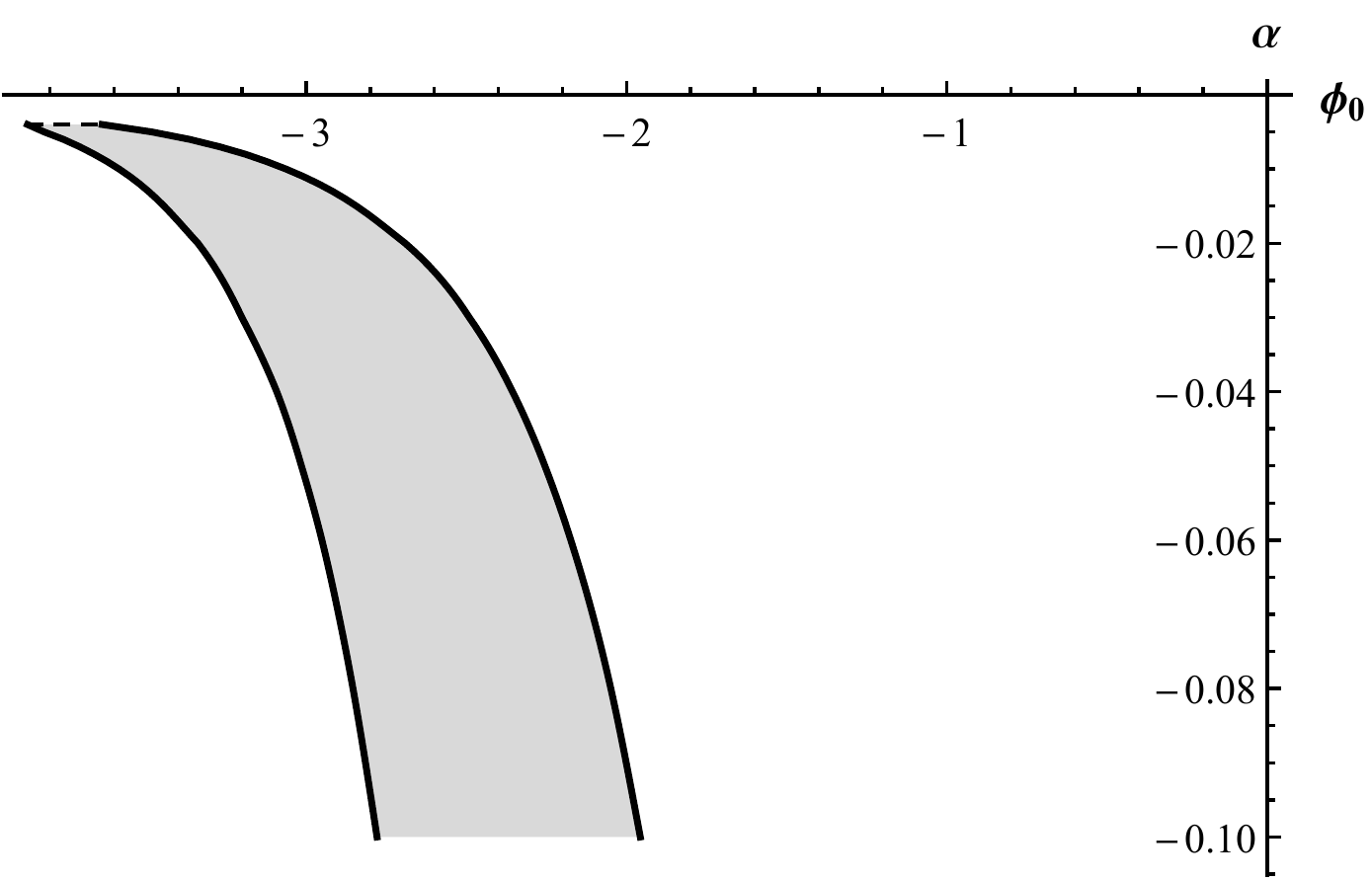}}
\caption{\label{fig:forbidden}Solutions are forbidden in gray regions. The parameters are fixed as $\kappa=0.1$, $\gamma=1.0$ and $U_0=\mp0.3$.}
\end{figure}

In the numerical computation to find the solutions, we noticed that there exists a region, in which the solution is forbidden under the specific value of $\phi_0$ and $\alpha$. For those parameters, the equations of motion seem to diverge in the computation. Fig.\ \ref{fig:forbidden} represent the forbidden region for AdS and dS background with respect to $\phi_0$ and $\alpha$. These regions only appear when $\alpha$ has the negative value. The forbidden region becomes bigger when $\phi_v$ decreases. In addition, the region is going to be far from zero when $\alpha$ increases. We cannot obtain the exact forbidden region in the vicinity of $\alpha=0$. Thus, there remain the upper bound with the dashed line. The forbidden region seems to appear due to the solution is imaginary. Similar things happen for the black holes in DEGB theory \cite{Ohta:2009tb}.

\begin{figure}[t]
\centering
\subfigure[\ $\phi(\eta)$ vs. $\eta$ for AdS background]{\includegraphics[width=0.45\textwidth,height=0.15\textheight]{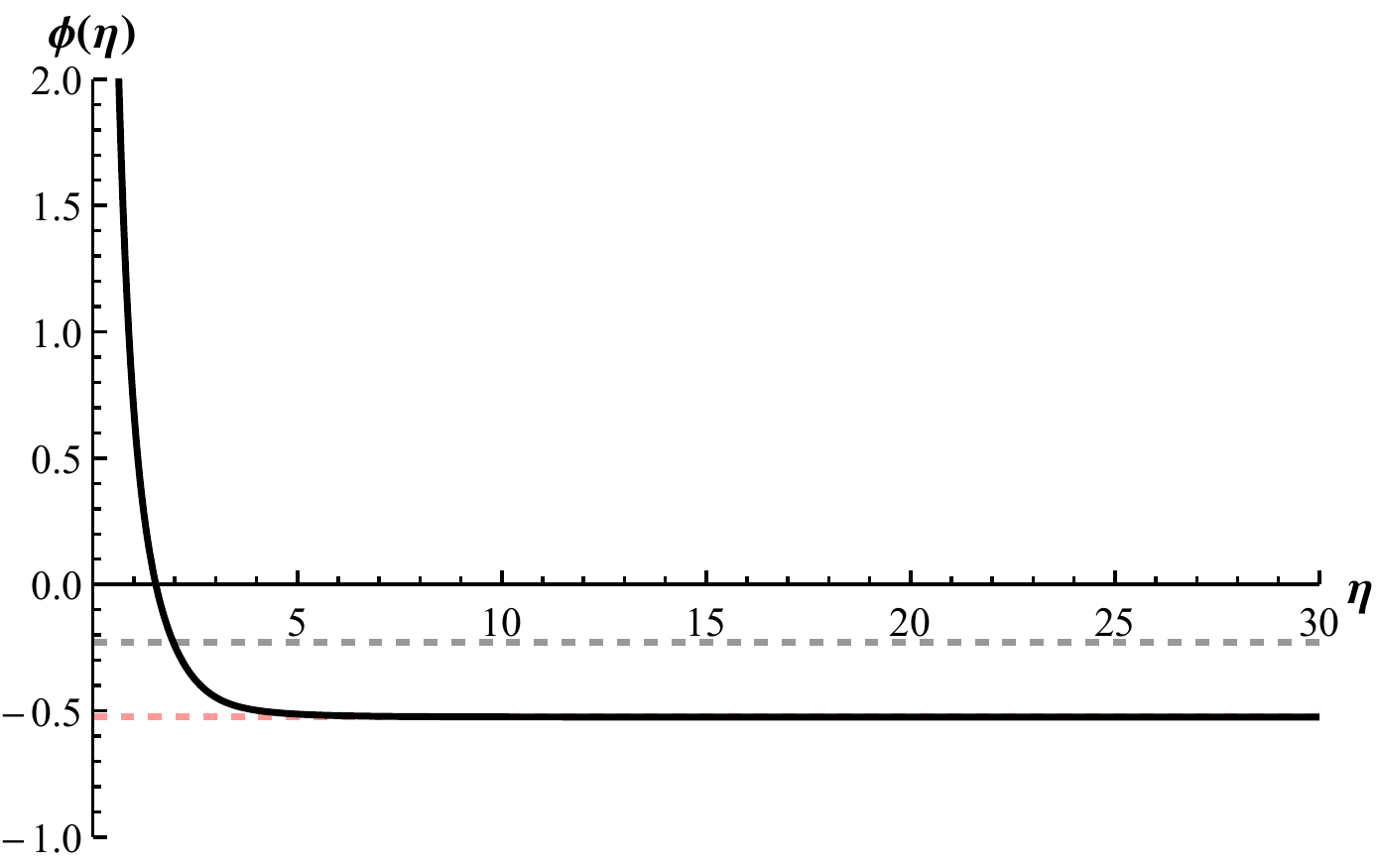}}
\hspace{0.03\textwidth}
\subfigure[\ $\rho(\eta)$ vs. $\eta$ for AdS background]{\includegraphics[width=0.45\textwidth,height=0.15\textheight]{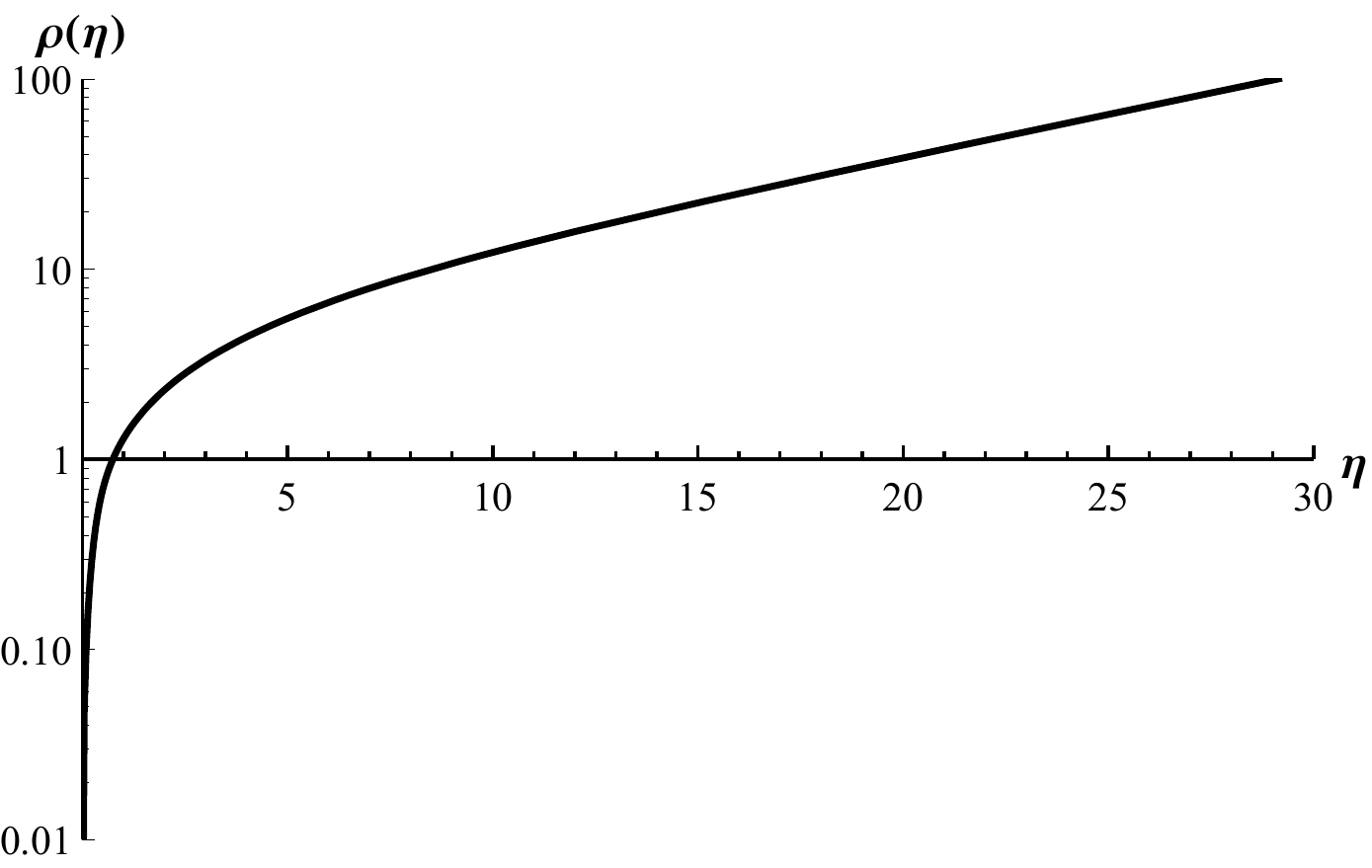}}
\caption{\label{fig:new.sol}New solution in the AdS background is appeared. The parameters are fixed as $\kappa = 0.1$, $\alpha = -0.1$, $\gamma = 8.0$ and $U_0= -0.3$.}
\end{figure}

\begin{figure}[ht]
\centering
\subfigure[\ $\phi(\eta)$ vs. $\eta$ for dS background]{\includegraphics[width=0.45\textwidth,height=0.15\textheight]{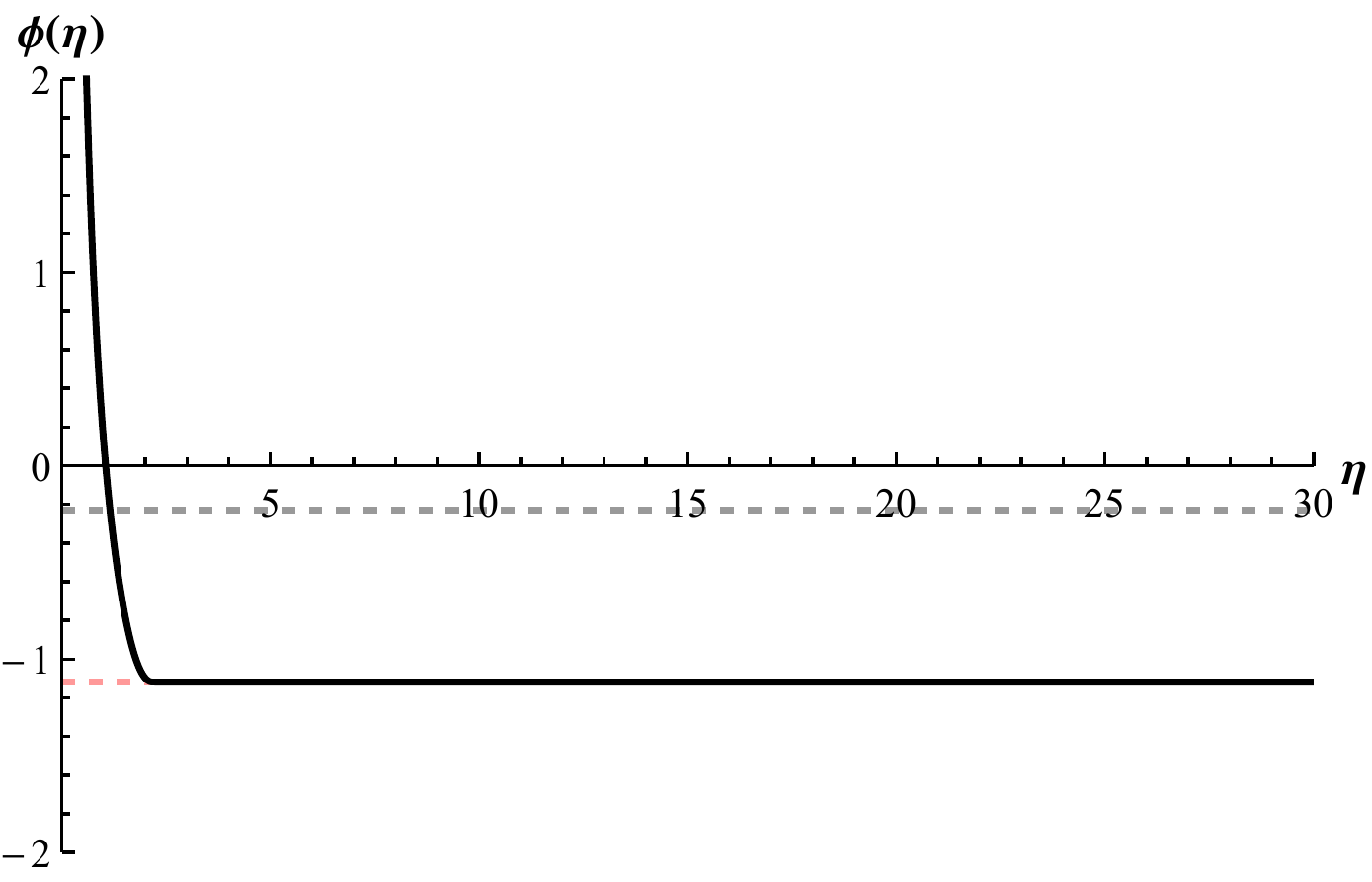}}
\hspace{0.03\textwidth}
\subfigure[\ $\rho(\eta)$ vs. $\eta$ for dS background]{\includegraphics[width=0.45\textwidth,height=0.15\textheight]{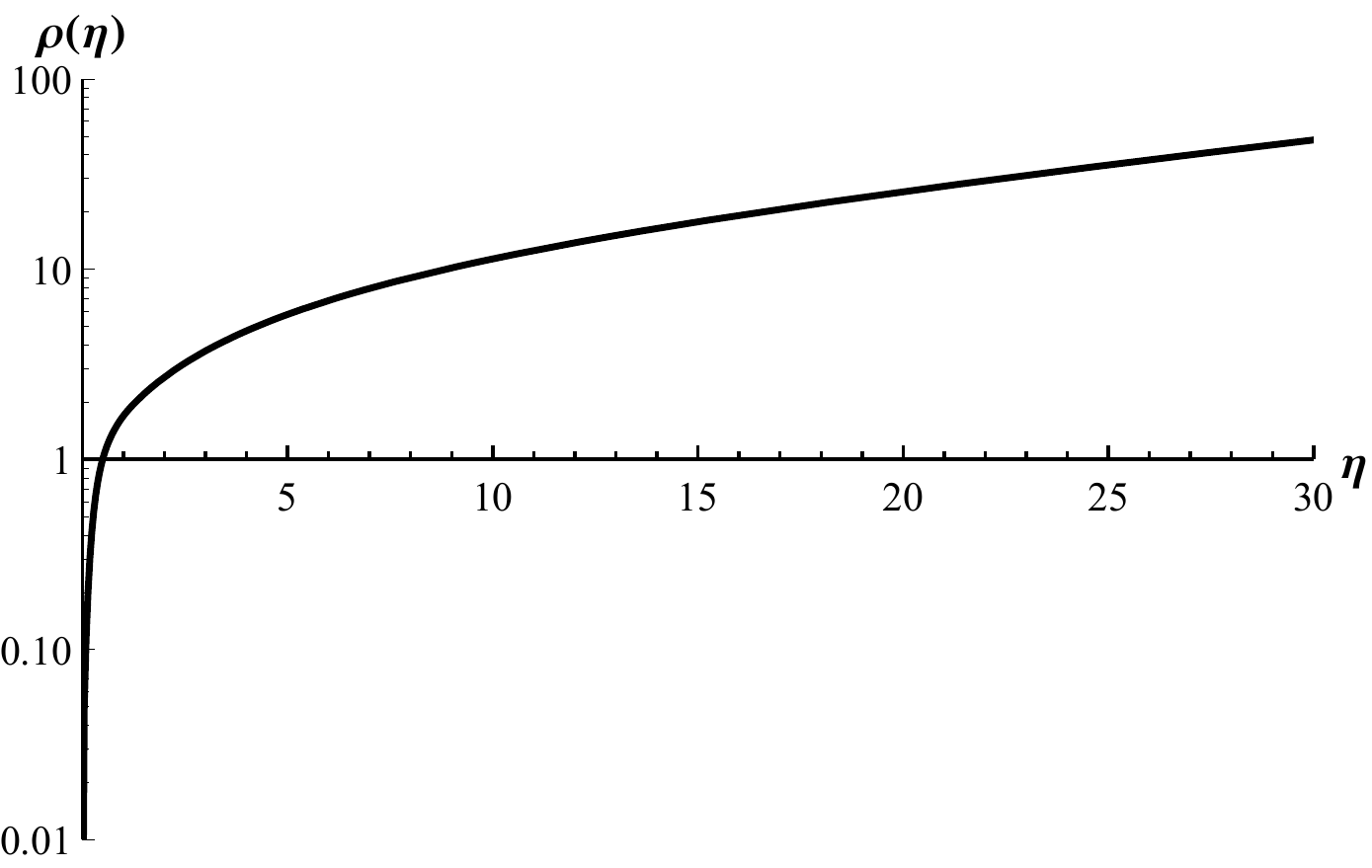}}
\caption{\label{fig:new.sol2}New solution in the dS background is appeared. The parameters are fixed as $\kappa = 0.1$, $\alpha = -0.1$, $\gamma = 8.0$ and $U_0= 0.3$.}
\end{figure}

Until now, we have only considered the solutions with $\gamma=1.0$ by changing $\alpha$, which solutions are already shown in Einstein theory. In order to find a new type of solution which is expected from the shape of the effective potential as in Fig.\ \ref{fig:alpha.vac}(b), we increase the gamma value into $\gamma=8.0$ and try to find solution. As a result, Figs.\ \ref{fig:new.sol} and \ref{fig:new.sol2} represent the new solution in the AdS and dS backgrounds. The black and red dashed lines indicate the first and second vacuum in both figures. In this new solution, the scalar field directly go to the second vacuum state and stop there, while the other solutions are converge or oscillate only for the first vacuum state.


\section{\label{sec:5}Summary and Discussion}
In this paper, we have investigated various types of Fubini instanton in DEGB theory. It is worthwhile to consider the hilltop potential in line with the recently preferred inflationary scenarios. In particular, we have taken into consideration the hilltop potential which is made only by the quartic scalar field term. The vacuum state is at $\phi=0$ in Einstein theory, $\phi_v$ in DEGB theory. Furthermore, the second vacuum state is formed when there is a negative coefficient $\alpha$. The existence of the second vacuum state affects the shape of the potential. Thus, the effective potential is changed to be bounded below as a result of the negative direction of potential which significantly changes the solution. This change becomes substantial when we set a higher value fir the dilaton coupling constant $\gamma$. For the consistent comparison of the solutions, we have fixed $\gamma = 1.0$.

We had investigated various types of solutions in Einstein theory in our previous paper \cite{Lee:2014ula}. In this paper, we presente solutions with respect to the GB coefficient $\alpha$ as an extension of our previous work. Since the vacuum state is shifted, the oscillation numbers of the solutions in the AdS background have changed. If the solution has a positive initial value, it displays a more oscillatory behavior when $\alpha$ increases and vice versa. In order to check the change in probability, we have calculated the exponent of the decay rate $B$ for the marginal solution, which only has a finite exponent $B$ in the AdS background \cite{Lee:2014ula}, with respect to $\alpha$. By doing so, the exponent $B$ decreases and increases again when $\alpha$ increases. In the other way around, the exponent $B$ decreases and increases again when $\alpha$ decreases. As a consequence of decreasing this exponent, it is possible to choose the more probable case by selecting a specific set of parameters. In the dS background, we only consider the $Z_2$-symmetric solution, and the symmetry is broken when $\alpha$ is introduced. We have also found that the initial values of the scalar field are restricted to be smaller than the second vacuum states, due to the face that the scalar field grows into negative infinity. Furthermore, the equations of motion cannot allow for regular behavior during numerical calculation for certain values of the initial scalar field. Thus, the solutions are forbidden for the specific parameter region for $\phi_0$ and $\alpha$. We have found a new solution using a second vacuum state which can be predicted by the form of the effective potential. The scalar field is neither oscillating nor converging to the specific value of scalar field, i.e.\ the first vacuum state, but instead exactly stops at the second vacuum state which is never shown in Einstein theory.

Recent inflation scenarios are focused on the hilltop or plateau inflationary models which are preferred for precision observations. In line with this, we have considered quantum tunneling on hilltop potential. We may compare the probability between rolling as an inflation scenario and tunneling as we have done in this paper. The timescale (or the decay probability) of tunneling on the potential may be larger (or smaller) than that of rolling in Einstein theory. However, we have found that tunneling probability can be improved by choosing the proper set of GB parameters in DEGB theory. By employing this method, the tunneling event is improved sufficiently and the effects can be seen in the observational data. In order to examine this point, we should consider the decay modes, i.e.\ rolling and tunneling simultaneously in a specific model. We hope to adopt our result into cosmology and consider this decay mode as well as the rolling mode in an inflationary scenario for future work.

\section*{Acknowledgments}
We would like to thank Seoktae Koh for his hospitality during our visit to Jeju National University. BHL was supported by National Research Foundation of Korea (NRF) grant funded by the Korea government (MSIP) (No.2014R1A2A1A01002306). WL was supported by Basic Science Research Program through National Research Foundation of Korea (NRF) funded by the Ministry of Education (2016R1D1A1B01010234). DH was supported by the Korea Ministry of Education, Science and Technology, Gyeongsangbuk-Do and Pohang City.

\bibliographystyle{jhep_style}
\section*{References}
\bibliography{References}
\end{document}